\documentclass[preprint,5p, times, twocolumn]{elsarticle}

\usepackage{amssymb}
\usepackage{booktabs}
\usepackage{multirow}
\usepackage{stfloats}
\usepackage[ruled,vlined]{algorithm2e}
\usepackage{hyperref}
\hypersetup{
	colorlinks=true,
}

\journal{arXiv.org}

\begin{document}

\begin{frontmatter}



\title{Real-time chaotic video encryption based on multithreaded\\parallel confusion and diffusion}


\author[label1,label2]{Dong Jiang\corref{cor}}
\cortext[cor]{Corresponding author.}
\ead{jiangd@nju.edu.cn}
\affiliation[label1]{
	organization={School of Internet},
    addressline={Anhui University}, 
    city={Hefei},
    postcode={230039}, 
    country={China}}
\affiliation[label2]{
	organization={National Engineering Research Center of Agro-Ecological Big Data Analysis and Application},
	addressline={Anhui University}, 
	city={Hefei},
	postcode={230601}, 
	country={China}}

\author[label1]{Zhen Yuan}
\author[label1]{Wen-xin Li}

\author[label3,label4,label5]{Liang-liang Lu\corref{cor}}
\affiliation[label3]{
	organization={Key Laboratory of Optoelectronic Technology of Jiangsu Province},
	addressline={Nanjing Normal University}, 
	city={Nanjing},
	postcode={210023}, 
	country={China}}
\affiliation[label4]{
	organization={School of Physical Science and Technology},
	addressline={Nanjing Normal University}, 
	city={Nanjing},
	postcode={210023}, 
	country={China}}
\affiliation[label5]{
	organization={National Laboratory of Solid State Microstructures},
	addressline={Nanjing University}, 
	city={Nanjing},
	postcode={210093}, 
	country={China}}
\ead{lianglianglu@nju.edu.cn}

\begin{abstract}
Due to the strong correlation between adjacent pixels, most image encryption schemes perform multiple rounds of confusion and diffusion to protect the image against attacks. Such operations, however, are time-consuming, cannot meet the real-time requirements of video encryption. Existing works, therefore, realize video encryption by simplifying the encryption process or encrypting specific parts of video frames, which results in lower security compared to image encryption.
To solve the problem, this paper proposes a real-time chaotic video encryption strategy based on multithreaded parallel confusion and diffusion. It takes a video as the input, splits the frame into subframes, creates a set of threads to simultaneously perform five rounds of confusion and diffusion operations on corresponding subframes, and efficiently outputs the encrypted frames.
The encryption speed evaluation shows that our method significantly improves the confusion and diffusion speed, realizes real-time $480\times 480$, $576\times576$, and $768\times768$ 24FPS video encryption using Intel Core i5-1135G7, Intel Core i7-8700, and Intel Xeon Gold 6226R, respectively. The statistical and security analysis prove that the deployed cryptosystems have outstanding statistical properties, can resist attacks, channel noise, and data loss.
Compared with previous works, to the best of our knowledge, the proposed strategy achieves the fastest encryption speed, and realizes the first real-time chaotic video encryption based on multi-round confusion-diffusion architecture, thus, providing a more secure and feasible solution for practical applications and related research.
\end{abstract}




\end{frontmatter}


\section{Introduction}

\begin{figure*}[b]
	\centering
	\includegraphics[width=0.84\textwidth]{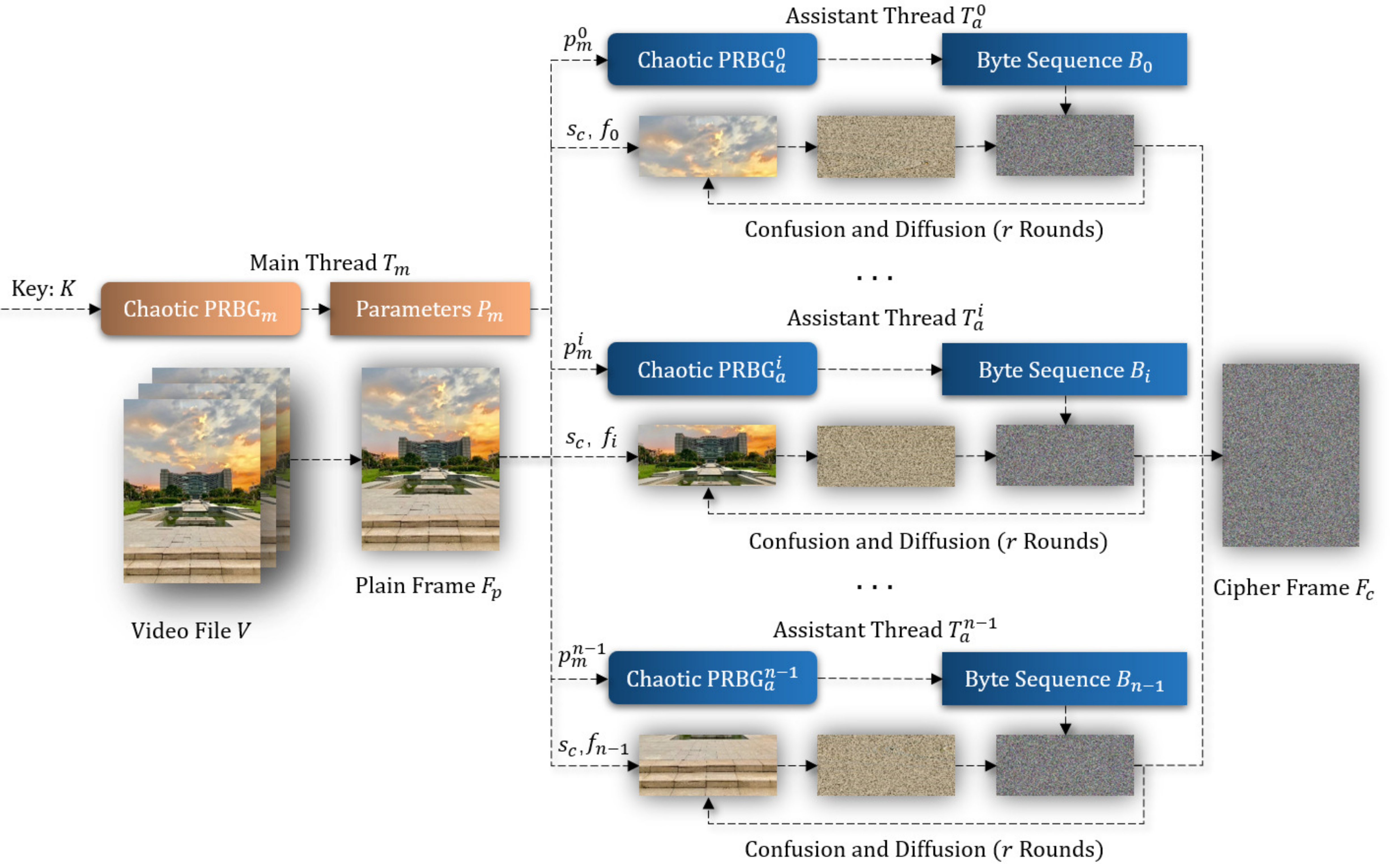}
	\caption{Workflow diagram of the proposed chaotic real-time video encryption strategy ($p_m$: parameters generated by $\mathrm{PRBG}_m$; $s_c$: confusion seed; $f$: subframe).}
	\label{Fig:StrategyDescription}
\end{figure*}

With the rapid development of information and communication technologies, images and videos have shown enormous potential in data storage and network transmission, resulting in extensive application requirements for image and video encryption \cite{pareek2006image}.
However, most conventional cryptographic schemes, such as DES, AES, RSA, etc., are designed to protect textual information, they are not suitable for images and videos \cite{fang2022survey}. As a result, many image encryption protocols have been proposed over recent years based on different techniques \cite{zhang2008optical, zhang2014symmetric, tedmori2014image, chai2017visually}, in which chaos based methods attract significant attention, due to the intrinsic characteristics of chaotic systems, including ergodicity, non-periodicity, non-convergence, sensitivity to initial conditions and control parameters, et al. \cite{luo2018parallel}. Most chaos based image encryption algorithms comprises confusion and diffusion phases \cite{zia2022survey}. In the former phase, the pixel positions are scrambled over whole image without changing the values \cite{chen2014fast}. In the latter phase, the pixel values are modified sequentially with the byte sequences generated by chaotic systems \cite{wong2009efficient}.

Such confusion-diffusion architecture based image encryption protocols need to perform the two phases for multiple rounds until a satisfactory security level is achieved \cite{matthews1989derivation}. This, obviously, is very time-consuming and cannot meet the real-time requirements of video encryption. Existing works, therefore, realize video encryption through simplifying encryption process or encrypting specific pixels in the video\cite{massoudi2008overview, liu2010survey}.
For the first category (also known as the full encryption), for example,
Ref. \cite{chiaraluce2002new} selects three chaotic maps to generate byte sequences, and directly performs XOR operations between the pixels and the generated bytes to encrypt the video;
Refs. \cite{lian2007chaotic} use the generated bytes to encrypt the frame, and take the encrypted pixels as feedback to improve the plaintext sensitivity of the deployed cryptosystem;
Ref. \cite{yang2008video} performs one round of confusion operation on the video frame, followed by acting XOR operations between the pixels and the generated bytes to realize frame encryption;
Since practical applications put forward higher requirements for security, most recently published works are based on one-round confusion-diffusion architecture\cite{valli2017chaos, ganeshkumar2019new, li2020video, yasser2020chaotic, hosny2022fast, dua20223d}.
The second category is also known as the selective encryption, the algorithms belong to this category encrypt specific pixels in the video frame to reduce the computational complexity \cite{hamidouche2017real}.
These categories of strategies, clearly,  achieve high efficiency at the expense of security.

Therefore, how to realize real-time video encryption without compromising the security has become an urgent problem to be solved. In the field of full video encryption, however, there are few related research. And to the best of our knowledge, existing works cannot realize real-time video encryption, i.e., the the number of frames encrypted in a second is larger than FPS (Frames Per Second) of the video or the average encryption time (ms) is less than $1000~/~$ FPS.
This paper, thus, takes the advantage of parallel computing, designs a five-round confusion-diffusion architecture based real-time chaotic video encryption strategy.
To evaluate the performance, two cryptosystems are implemented using two different chaotic maps. Three hardware platforms are used to assess the encryption speed of the deployed cryptosystems. The evaluation results show that our strategy significantly improves the speed of byte generation, confusion, and diffusion, laying the foundation for the realization of real-time video encryption. The statistical and security analysis prove that the deployed cryptosystems have outstanding statistical properties and resistance to attacks,  channel noise and data loss. The proposed strategy also suitable for many confusion and diffusion methods, and can be easily realized with both software and hardware.

The rest of the paper are organized as follows:
section 2 gives a detailed description of the proposed strategy.
In section 3, two typical chaotic maps are selected to realize the strategy, and the encryption speed of the deployed cryptosystems are evaluated using three different hardware platforms.
Section 4 and 5 carry out statistical and security analysis, respectively.
Section 6 analyzes the reasons for the parameter settings used in this paper.
In section 7 gives a comparison to recent published works, followed by a brief conclusion in section 8.

\section{Strategy Description}
In the proposed strategy, as shown in Fig. \ref{Fig:StrategyDescription}, a set of threads are used to encrypt the video frames, simultaneously. These threads can be divided into two categories: a main thread $T_m$ and $n$ assistant threads $T_a^i~(i\in\{0,...,n-1\})$. The main thread takes the key $K$ and the original video $V$ as input, initializes its $\mathrm{PRBG}_m$, generates the parameters $P_m$, creates and organizes $T_a$s to perform confusion and diffusion operations on the video frames. See Algo. \ref{alg:mainThread} for the detailed encryption process of  $T_m$.

\begin{algorithm}[!h]
	\caption{Encryption process of main thread $T_m$.}
	\label{alg:mainThread}
	\KwIn{Key: $K$; Number of assistant threads: $n$; Video file: $V$; Rounds of confusion and diffusion operations: $r$. }
	\KwOut{Parameters for initializing $\mathrm{PRBG}_a$s of $T_a$s: $P_m$; Plain Frame: $F_p$; Confusion seed: $s_c$.}
	\BlankLine
	Initialize $\mathrm{PRBG}_m$ with $K$;\\
	Iterate $\mathrm{PRBG}_m$, generate enough iteration results;\\
	Convert iteration results to parameters $P_m$;\\
	\For{$i = 0; i < n; i++$}{
		Create the assistant thread $T_a^i$;\\
	}
	\While{$\mathrm{Fetch~a~frame}~F_p~\mathrm{from}~V$}{
		Iterate $\mathrm{PRBG}_m$, and generate a confusion seed $s_c$;\\
		\For{$i = 0; i < r; i++$}{
			Awake $T_a$s to perform confusion operation;\\
			Wait for $T_a$s to finish the confusion operation;\\
			Awake $T_a$s to perform diffusion operation;\\
			Wait for $T_a$s to finish the diffusion operation;\\
		}
	}
\end{algorithm}

Assistant thread $T_a^i$, initializes its $\mathrm{PRBG}_a^i$ with parameters $p_m^i\in P_m$, generates iteration results, converts the results into byte sequence $B_i$, performs confusion and diffusion operations on the subframe $f_i$, and outputs the cipher frame $F_c$.
In confusion operation, a discretized version of the Chirikov normal map \cite{chen2018optical} as follows is used to rearrange the pixel positions:

\begin{equation}
	\left\{
	\begin{array}{lr}
		\alpha=(a + o)~\mathrm{mod} ~w,\\
		\beta=(o + s_c\mathrm{sin}(2\pi \alpha / w))~\mathrm{mod}~w,\\
	\end{array}
	\right.
	\label{equ:ChirikovStandardMap}
\end{equation}
where $a$ and $o$ are the abscissa and ordinate of the pixel, $\alpha$ and $\beta$ are the new coordinates of the pixel after confusion operation, $w$ is the width of the video frame, and $s_c$ is referred to as the confusion seed, which is a positive integer generated by $T_m$.
The inverse transform of confusion operation is as follows:
\begin{equation}
	\left\{
	\begin{array}{lr}
		a=(\alpha - \beta + s_c\mathrm{sin}(2\pi \alpha / w))~\mathrm{mod}~w,\\
		o=(\beta - s_c\mathrm{sin}(2\pi\alpha / w)) ~\mathrm{mod}~w.\\
	\end{array}
	\right.
	\label{equ:InverseChirikovStandardMap}
\end{equation}

\begin{algorithm}[h]
	\caption{Encryption process of assistant thread $T_a^i$.}
	\label{alg:assistantThread}
	\KwIn{Thread index: $i$; Number of assistant threads: $n$; Plain frame: $F_p$; Width of $F_p$: $w$; Parameters generated by $T_m$: $P_m$; Rounds of confusion and diffusion $r$; Confusion seed: $s_c$.}
	\KwOut{Cipher frame: $F_c$.}
	\BlankLine
	startRow = $i\times w/n$;\\
	endRow = startRow $+ w/n$;\\
	Initialize $\mathrm{PRBG}_a^i$ with $p_m^i$;\\
	\While{$\mathrm{Fetch~subframe}~f_i~\mathrm{from}~F_p$}{
		Iterate $\mathrm{PRBG}_a^i$, generate enough iteration results;\\
		Convert iteration results into byte sequence $B_i$;\\
		$F_c=F_p$;\\
		\For{$j=0; j<r; j++$}{
			Wait to be woken up by $T_m$;\\
			Temp frame $F_t = F_c$;\\
			\For{$a = \mathrm{startRow}; a < \mathrm{endRow}; a++$}{
				\For{$o = 0; o < w; o++$} {
					$\alpha = (a + o)~\mathrm{mod}~w$;\\
					$\beta = (o + s_c\mathrm{sin}(2\pi \alpha / w))~\mathrm{mod}~w$;\\
					$F_c[\alpha,\beta] = F_t[a,o]$;\\
				}
			}
			Tell $T_m$ that the confusion operation is completed\\
			Wait to be woken up by $T_m$\\
			$F_t = F_c$;\\
			Take the last pixel of $f_{(i+1)\mathrm{mod}~n}$ as diffusion seed $s_d$;\\
			\For{$a = \mathrm{startRow}; a < \mathrm{endRow}; a++$}{
				\For{$o = 0; o < w; o++$} {
					Fetch a byte $b$ from $B_i$;\\
					\eIf{$a==\mathrm{startRow~AND}~o==0$}{
						$F_c[a,o]=b\oplus(F_t[a,o]+b)\oplus s_d$;\\
					}{
						Fetch previous pixel $F_c[a',o']$;\\
						$F_c[a,o]=b\oplus(F_t[a,o]+b)\oplus F_c[a',o']$;\\
					}
				}
			}
			Tell $T_m$ that the diffusion operation is completed\\
		}
	}
\end{algorithm}

In diffusion, the following equation is employed\cite{fu2012chaos}:

\begin{equation}
	F_c[a,o]=b\oplus (F_p[a,o] + b)\oplus F_c[a',o'],
	\label{Equ:diffusion}
\end{equation}
where $F_c[a,o]$ and $F_p[a, o]$ denote the encrypted and the plain pixel, $b$ is a byte generated by $\mathrm{PRBG}_a^i$, and $[a', o']$ is the coordinate of previous pixel.
To perform diffusion operation, the diffusion seed $s_d$ needs to be set to encrypt $F_p[0,0]$. In the proposed strategy, $T_a^i$ takes the last pixel of subframe $f_{(i+1)~\mathrm{mod}~n}$ as $s_d$. The inverse transform of diffusion is as follows:
\begin{equation}
	F_p[a,o]=(b \oplus F_c[a,o]\oplus F_c[a', o']) - b.
	\label{Equ:inverseDiffusion}
\end{equation}
See Algo. \ref{alg:assistantThread} for the detailed encryption process of $T_a^i$.

\section{Encryption Speed Evaluation}
To evaluate the encryption speed of the proposed strategy, two highly-cited chaotic maps are selected to implement PRBGs. The first PRBG uses two Piecewise Linear Chaotic Maps (PLCM) \cite{zhong2018self}, which is defined as follows:
\begin{equation}
	x_{i+1}=F(x_i, p)=
	\left\{
	\begin{array}{lr}
		x_i/p, ~~~~~~~~~~~~~~~~~~~~x_i\in[0,p)\\
		(x_i-p)/(\frac{1}{2}-p), ~x_i\in[p,0.5],\\
		F(1-x_i,p), ~~~~~~~~x_i\in(0.5,1]
	\end{array}
	\right.
	\label{equ:plcm}
\end{equation}
where $x_i\in[0,1]$, $x_0$ and $p\in(0,0.5)$ are referred to as the initial condition and the control parameter, respectively. The second PRBG employs couple two dimensional Logistic-Adjusted-Sine Map (2DLASM) \cite{hua2016image}, which is given by:
\begin{equation}
	\left\{
	\begin{array}{lr}
		x_{i+1}=\mathrm{sin}(\pi\mu(y_i +3)x_i(1-x_i)),\\
		y_{i+1}=\mathrm{sin}(\pi\mu(x_{i+1}+3)y_i(1-y_i)),\\
	\end{array}
	\right.
	\label{equ:2DLASM}
\end{equation}
where $x_i, y_i\in[0,1]$, $x_0$ and $\mu\in[0.37, 0.38]\cup[0.4, 0.42]\cup[0.44,0.93]\cup\{1\}$ are referred to as the initial condition and the control parameter, respectively.

Three hardware platforms are selected to carry out the evaluations. The specifications of the platforms are listed in Tab. \ref{Tab:PlatformSetup}. The software development environment are as follows: Ubuntu 20.04 operating system, OpenCV 4.2.0, and g++ 9.4.0.

\begin{table}[!thpb]\small
	\caption{The specifications of hardware platforms}
	\begin{tabular}{p{1.8cm}p{2.8cm}p{3.0cm}}
		\toprule
		Platform                                & Specifications         & \\
		\midrule
		\multirow{4}{1cm}{Laptop}
		& CPU               & Intel Core i5-1135G7\\
		& Cores / Threads   & 4 / 8\\
		& Frequency         & 2.4GHz\\
		& Memory            & 8GB\\
		\midrule
		\multirow{4}{1cm}{Personal Computer}
		& CPU               & Intel Core i7-8700\\
		& Cores / Threads   & 6 / 12\\
		& Frequency         & 3.2GHz\\
		& Memory            & 32GB\\
		\midrule
		\multirow{4}{1cm}{Workstation}
		& CPU               & Intel Xeon Gold 6226R\\
		& Cores / Threads   & 16 / 32\\
		& Frequency         & 2.9GHz\\
		& Memory            & 64GB\\
		\bottomrule
		\label{Tab:PlatformSetup}
	\end{tabular}
\end{table}

\begin{figure*}[b]
	\centering
	\includegraphics[width=0.98\textwidth]{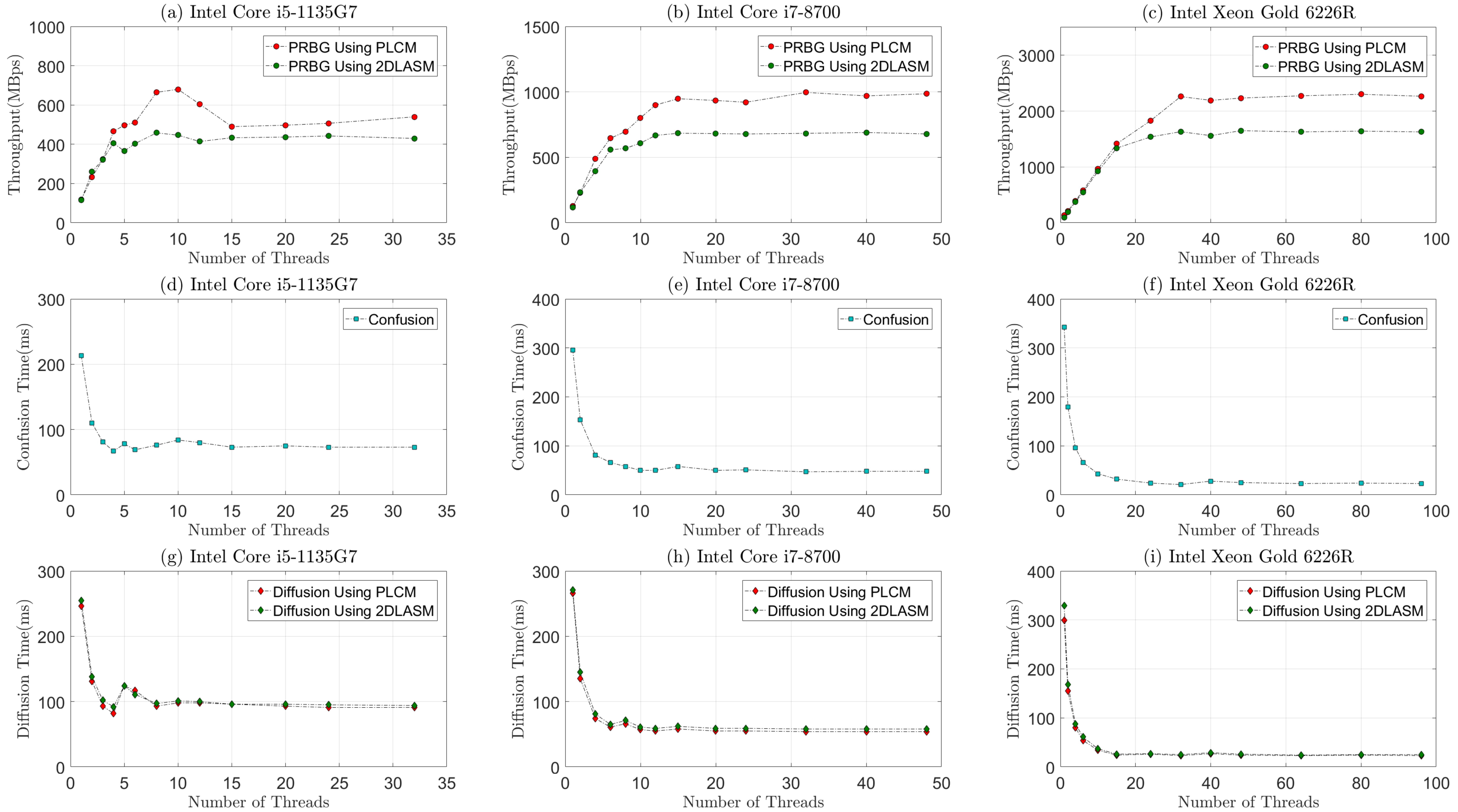}
	\caption{Speed evaluation of byte generation, confusion, and diffusion steps, (a-c) throughput of all $\mathrm{PRBG_a}$s versus the number of assistant threads, (d-f) average time of confusion operations versus the number of assistant threads, (g-i) average time of diffusion operations versus the number of assistant threads.}
	\label{Fig:throughputEvaluation}
\end{figure*}

Since the encryption process of the proposed strategy consists of byte generation, confusion, and diffusion, we implement two cryptosystems using the selected chaotic maps, separately evaluate the speed of these steps, and assess the video encryption speed of the deployed cryptosystesm. In these experiments, the keys $K$ used to initialize PRBG$_m$ are randomly selected, the original video is downloaded from the \href{http://trace.eas.asu.edu/yuv/index.html}{video trace library}, and is converted into different sizes for encryption speed evaluation.

First, different number of assistant threads are used to generate byte sequences, simultaneously.
In PLCM based cryptosystem, the assistant thread $T_a^i$ uses $p_m^i$ generated by the main thread $T_m$ to initialize its $\mathrm{PRBG}_a^i$, iterates PLCMs and generates two sets of iteration results $\{x_1^1,x_2^1,x_3^1...\}$, $\{x_1^2,x_2^2,x_3^2...\}$, fetches $6$ bytes from the mantissa part\footnote{In IEEE 754 standard, a double precision floating-point number consists of 1 bit sign, 11 bits exponent, and 52 bits mantissa fields.} of each iteration result, and produces two sets of byte sequences $\{b_1^1,b_2^1,b_3^1,...\}$, $\{b_1^2,b_2^2,b_3^2,...\}$. By sequentially performing XOR operations on the generated bytes, $T_a^i$ outputs the final byte sequence $B_i$, which can be used for video encryption and decryption.
In 2DLASM based cryptosystem, similarly, $T_a^i$ uses $p_m^i$ to initialize its $\mathrm{PRBG}_a^i$, iterates the 2DLASMs and generates two sets of iteration results $\{x_1^1,y_1^1,x_2^1,y_2^1,...\}$, $\{x_1^2,y_1^2,x_2^2,y_2^2,...\}$. By fetching $6$ bytes from each iterations result and acting XOR operations on the generated bytes, $T_a^i$ produces the byte sequence $B_i$ . To calculate the throughput, all chaotic maps are iterated for $5\times 10^7$ times. The relationship between the throughput (MBps) and the number of assistant threads are shown in Fig. \ref{Fig:throughputEvaluation} (a)-(c).

Second, we set the rounds of confusion and diffusion $r$ to $5$ (see section 6 for the reasons), perform five rounds of confusion operations on 100 images of size $960\times 960$ with different number of assistant threads, calculate the average time to scramble an image, and plot the results in \ref{Fig:throughputEvaluation} (d)-(f).
Similarly, we use different number of assistant threads to generate byte sequences, perform five rounds of diffusion operations on 100 images of size $960\times 960$ using the generated bytes, and calculate the average time to complete five rounds of diffusion operations on an image.
The relationship between the average diffusion time (including byte generations and diffusion operations) are drawn in Fig. \ref{Fig:throughputEvaluation} (g)-(i).
Compared with the single-threaded versions of byte generation, confusion, and diffusion, clearly, the proposed strategy significantly speeds up these steps.

Third, to evaluate the practical performance of the proposed strategy, a set of videos of different sizes are encrypted using the deployed cryptosystems.
Similar to above experiments, the rounds of confusion and diffusion $r$ is set to $5$. The number of assistant threads $n$ are set to $8$, $12$, and $32$ for laptop, personal computer, and workstation, respectively (see section 6 for the reasons).
The Frames Per Second (FPS) of the selected vidoes is $20$, that is, the maximum time for encrypting a frame must be less than $50$ms, otherwise, there will exist delay during encryption and transmission. Each video consists of $600$ frames. The average encryption time are listed in Tab. \ref{Tab:videoEncryption}, which shows that real-time $576\times 576$, $672\times 672$, and $960\times 960$ video encryption are realized using Intel Core i5-1135G7, Intel Core i7-8700, and Intel Xeon Gold 6226R, respectively. 
\begin{figure*}[b]
	\centering
	\includegraphics[width=0.98\textwidth]{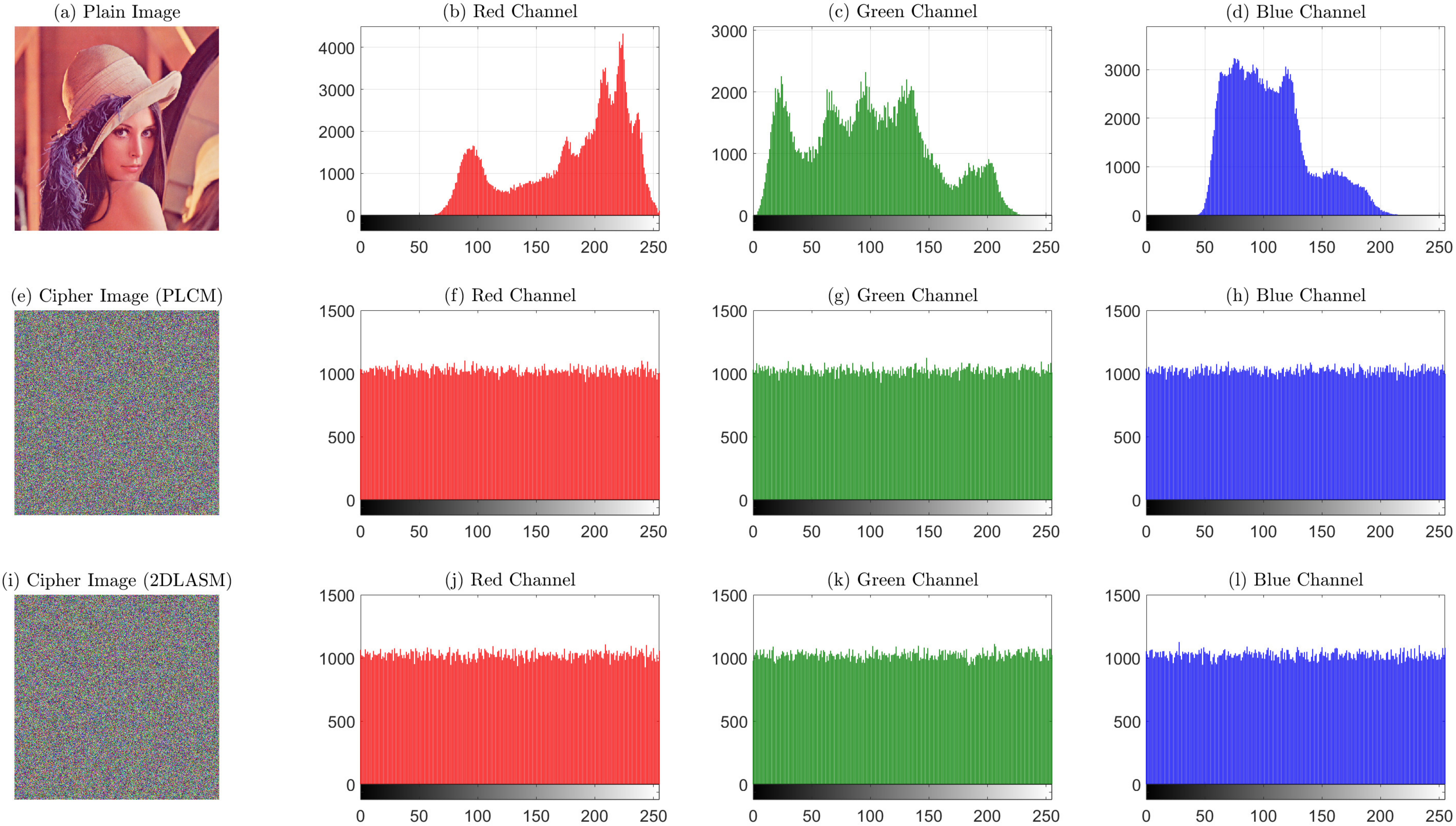}
	\caption{Histograms of plain and cipher images. (a) plain image Lena, (b) - (d) histograms of the red, green, and blue channels of the plain image, (e) cipher image encrypted with PLCM, (f) - (h) histograms of the red, green, and blue channels of the cipher image encrypted with PLCM, (i) cipher image encrypted with 2DLASM, (h) - (j) histograms of the red, green, and blue channels of the cipher image encrypted with 2DLASM.}
	\label{Fig:uniformityEvaluation}
\end{figure*}

\begin{table*}[!thbp]\small
\caption{Video encryption speed evaluation.}
\begin{tabular}{cc|cc|cc|cc}
\toprule
\multirow{3}{*}{Video Size} & \multirow{3}{*}{Max Encryption Time} &\multicolumn{2}{c}{Laptop (8 $T_a$s)} & \multicolumn{2}{|c|}{Personal Computer (12 $T_a$s)} & \multicolumn{2}{c}{Workstation (32 $T_a$s)}\\
& & \multicolumn{2}{c}{Average Encryption Time(ms)} & \multicolumn{2}{|c|}{Average Encryption Time(ms)} & \multicolumn{2}{c}{Average Encryption Time(ms)}\\
\cmidrule(r){3-8}
				  &      & PLCM     & 2DLASM   & PLCM      & 2DLASM    & PLCM      & 2DLASM\\
\midrule
$96\times 96$     & 42ms & 5.69     & 5.54     & 7.43      & 7.40      & 2.92      & 2.62\\
$192\times 192$   & 42ms & 16.59    & 16.56    & 21.45     & 20.92     & 7.69      & 7.41\\
$288\times 288$   & 42ms & 29.63    & 28.98    & 30.39     & 31.76     & 15.12     & 14.94\\
$384\times 384$   & 42ms & 31.60    & 30.64    & 35.97     & 35.22     & 23.25     & 22.94\\
$480\times 480$   & 42ms & 36.26    & 35.49    & 37.76     & 37.88     & 28.76     & 28.11\\
$576\times 576$   & 42ms & $\times$ & $\times$ & 39.92     & 40.67     & 30.40     & 30.02\\
$672\times 672$   & 42ms & $\times$ & $\times$ & $\times$  & $\times$  & 33.60     & 32.89\\
$768\times 768$   & 42ms & $\times$ & $\times$ & $\times$  & $\times$  & 36.56     & 36.23\\
\midrule
\multicolumn{8}{l}{ Number of Frames: 300; FPS: 24; Rounds of confusion and diffusion: $r=5$; $\times$: Average Encryption Time $\geq$ Max Encryption Time.}\\
\bottomrule
\label{Tab:videoEncryption}
\end{tabular}
\end{table*}

\section{Statistical Evaluation}
The essence of real-time video encryption is image encryption, which should provide excellent statistical properties to resist different types of attacks \cite{li2018cryptanalysis}.
In this section, therefore, the uniformity, correlation, and randomness of the deployed cryptosystems are analyzed. The following experiments are performed using the laptop (Intel Core i5-1135G7), and the parameter setup are as follows: number of assistant threads: $n = 8$, confusion and diffusion rounds $r=5$, the size of all plain images used for statistical evaluation is $512\times512$.

\subsection{Uniformity Evaluation}
The histogram, variances and $\chi^2$ of histograms are utilized to evaluate the uniformity of the deployed cryptosystems.
The plain image is plotted in Fig. \ref{Fig:uniformityEvaluation} (a), and its histograms of red, green, blue channels are shown in \ref{Fig:uniformityEvaluation} (b), (c), (d), respectively. The cipher image encrypted with PLCM is shown in Fig. \ref{Fig:uniformityEvaluation} (e). Its histograms of red, green, blue channels are shown in Fig. \ref{Fig:uniformityEvaluation} (f), (g), (h), respectively. The cipher image encrypted with 2DLASM is shown in Fig. \ref{Fig:uniformityEvaluation} (f). Its histograms of red, green, blue channels are shown in Fig. \ref{Fig:uniformityEvaluation} (j), (k), (l), respectively.

The variance of an image can be calculated as follows \cite{man2021double}:
\begin{equation}
	\mathrm{var(Z)} = \frac{1}{256^2}\sum_{i=0}^{255}\sum_{j=0}^{255}\frac{1}{2}(z_i-z_j)^2,
\end{equation}
where $\mathrm{Z} = \{z_0, z_1,...,z_{255}\}$ is the vector of the histogram values, $z_i$ and $z_j$ are the numbers of pixels whose values are equal to $i$ and $j$, respectively. The $\chi^2$ of an image is given by \cite{kwok2007fast}:
\begin{equation}
	\chi^2=\sum_{i=0}^{255}\frac{(z_i-N/256)^2}{N/256},
\end{equation}

\begin{table}[!thpb]\small
\caption{Variances and $\chi^2$ results of plain and cipher images}
\begin{tabular}{cl|ccc}
\toprule
\multirow{2}{0.6cm}{Test} 	    & \multirow{2}{2.6cm}{Image}  & \multicolumn{3}{c}{Channel}\\
\cmidrule(r){3-5}
							    &           		  & R         & G         & B\\
\midrule
\multirow{18}{0.6cm}{var}     	& Lena      		  & 1021324   & 457505    & 1382757\\
								& Lena (PLCM)         & 920.81    & 982.86    & 802.76\\
								& Lena (2DLASM)       & 1078.06   & 1047.92   & 1065.47\\
\cmidrule(r){2-5}
								& Airplane            & 2724339   & 2740687   & 4448810\\
								& Airplane (PLCM)     & 1139.74   & 1043.63   & 930.89\\
								& Airplane (2DLASM)   & 1022.57   & 1115.45   & 1035.92\\
\cmidrule(r){2-5}
								& Lake                & 789874    & 522660    & 1383691\\
								& Lake (PLCM)         & 942.18    & 1019.14   & 1166.58\\
								& Lake (2DLASM)       & 1063.56   & 913.51    & 1068.22\\
\cmidrule(r){2-5}
								& Peppers             & 856092    & 1278525   & 1973421\\
								& Peppers (PLCM)      & 981.84    & 995.96    & 919.01\\
								& Peppers (2DLASM)    & 1030.07   & 946.35    & 1101.70\\
\cmidrule(r){2-5}
								& Mandrill            & 332658    & 573472    & 321024\\
								& Mandrill (PLCM)     & 1018.60   & 1016.43   & 1138.86\\
								& Mandrill (2DLASM)   & 1165.99   & 993.28    & 872.41\\
\cmidrule(r){2-5}
								& Splash              & 2432151   & 3095990   & 5940168\\
								& Splash (PLCM)       & 1015.49   & 1112.96   & 1072.76\\
								& Splash (2DLASM)     & 954.40    & 880.06    & 1127.92\\
\midrule
\multirow{18}{0.6cm}{$\chi^2$}  & Lena                & 254333    & 113929    & 344338\\
								& Lena (PLCM)         & 229.30    & 244.75    & 199.90\\
								& Lena (2DLASM)       & 268.46    & 260.95    & 265.32\\
\cmidrule(r){2-5}
								& Airplane            & 678424    & 682495    & 1107858\\
								& Airplane (PLCM)     & 283.82    & 259.88    & 231.81\\
								& Airplane (2DLASM)   & 254.64    & 277.77    & 257.96\\
\cmidrule(r){2-5}
								& Lake                & 196697    & 130154    & 344571\\
								& Lake (PLCM)         & 234.62    & 253.79    & 290.50\\
								& Lake (2DLASM)       & 264.85    & 227.48    & 266.01\\
\cmidrule(r){2-5}
								& Peppers             & 213187    & 318382    & 491428\\
								& Peppers (PLCM)      & 244.50    & 248.01    & 228.85\\
								& Peppers (2DLASM)    & 256.51    & 235.66    & 274.34\\
\cmidrule(r){2-5}
								& Mandrill            & 82839     & 142808    & 79942\\
								& Mandrill (PLCM)     & 253.65    & 253.11    & 283.60\\
								& Mandrill (2DLASM)   & 290.35    & 247.35    & 217.25\\
\cmidrule(r){2-5}
								& Splash              & 605662    & 770974    & 1479241\\
								& Splash (PLCM)       & 252.88    & 277.15    & 267.14 \\
								& Spalsh (2DLASM)     & 237.66    & 219.15    & 280.88\\
\bottomrule
\label{Tab:uniformityEvaluation}
\end{tabular}
\end{table}

\noindent where $N$ is the number of pixels, $N/256$ is the expected occurrence frequency. The higher variance indicates the lower uniformity of an image, and when the significant level is $0.05$, $\chi^2(0.05, 255)$ is equal to 293.25 \cite{zhang2014chaotic}.
We encrypt a set of images using PLCM and 2DLASM, respectively, calculate and list the variance and $\chi^2$ test results of plain and cipher images in Tab. \ref{Tab:uniformityEvaluation}. Clearly, the variances of the cipher images are significantly reduced compared to the variances of the corresponding plain images, and $\chi^2$ values of the cipher images are lower than $293.25$. The deployed cryptosystems, therefore, uniformly conceal the information of the plain images.

\begin{figure*}[!thpb]
	\centering
	\includegraphics[width=0.98\textwidth]{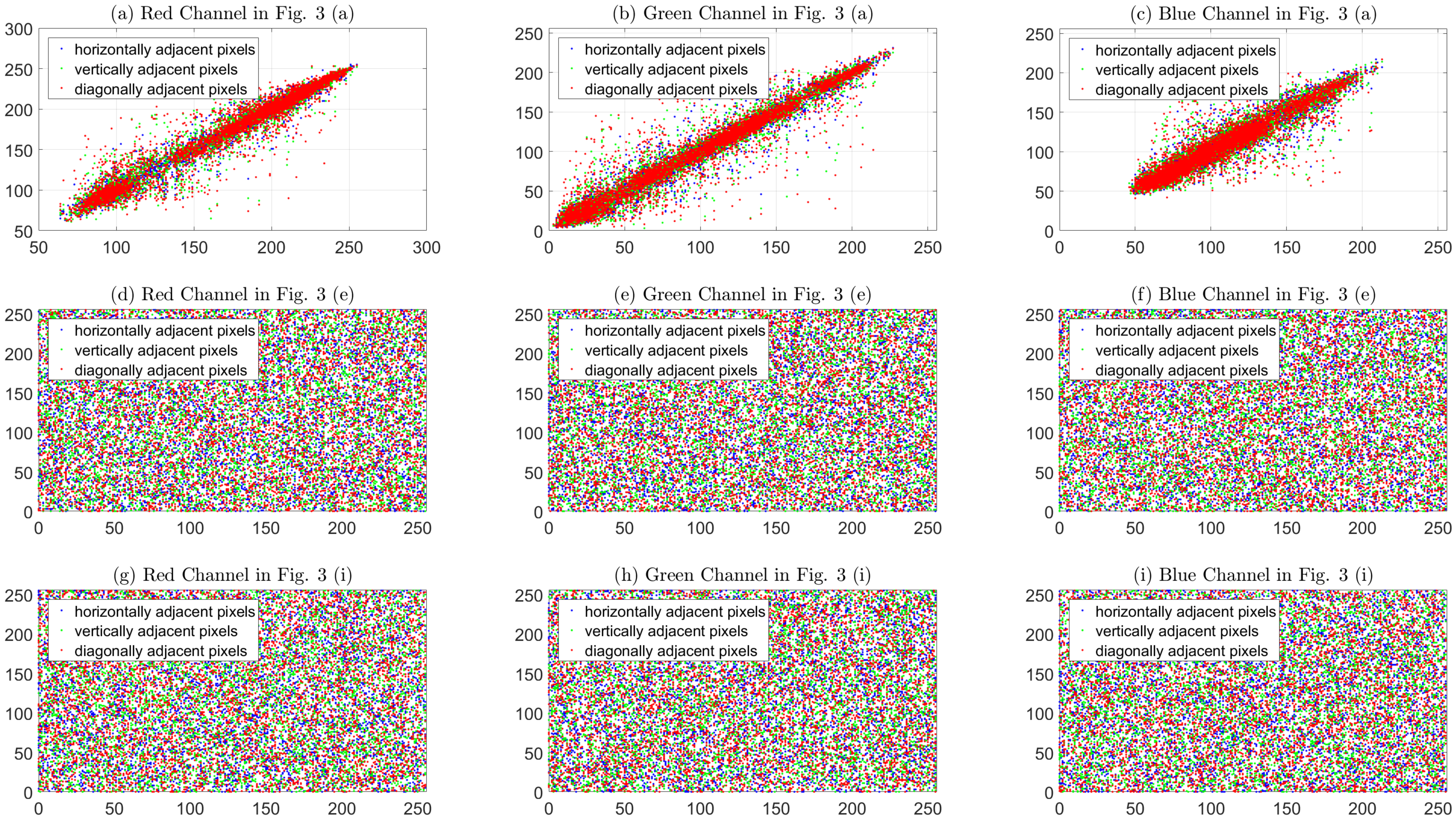}
	\caption{Correlation distribution of two adjacent pixels in horizontal, vertical, and diagonal directions, (a) - (c) correlation distribution of red, green, and blue channels in plain image, (d) - (f) correlation distribution of red, green, and blue channels in cipher image encrypted with PLCM, (g) - (i) correlation distribution of red, green, and blue channels in cipher image encrypted with 2DLASM.}
	\label{Fig:correlationEvaluation}
\end{figure*}

\begin{table*}[!thpb]\small
	\caption{Correlation coefficients of two adjacent pixels in the plain and cipher images}
	\begin{tabular}{lc|ccc|ccc|ccc}
		\toprule
		\multirow{2}{*}{Image} & \multirow{2}{*}{Channel} & \multicolumn{3}{c|}{Plain Image} & \multicolumn{3}{c|}{Cipher Image Encrypted With PLCM} & \multicolumn{3}{c}{Cipher Image Encrypted With 2DLASM}\\
		\cmidrule(r){3-11}
		& & H & V & D & H & V & D & H & V & D \\
		\midrule
		\multirow{3}{*}{Lena}      & R & 0.988469 & 0.980273 & 0.968900 & -0.003864 & - 0.000723 & -0.009886 & -0.001331 & -0.003863 & -0.002300   \\
		& G & 0.980549 & 0.968197 & 0.951883 & 0.013827  & 0.003009   & 0.002088  & -0.008544 & -0.002600 & -0.012616   \\
		& B & 0.951984 & 0.927028 & 0.908143 & 0.003075  & 0.002843   & 0.003700  & -0.002474 & 0.001525  & -0.016043   \\
		\midrule
		\multirow{3}{*}{Airplane} & R & 0.970642 & 0.970695 & 0.945469 & 0.006467  & -0.006111 & 0.003168  & 0.001575   & 0.011213  & -0.008233   \\
		& G & 0.971925 & 0.926737 & 0.902899 & -0.004944 & 0.003830  & -0.017420 & -0.003714  & 0.001359  & -0.008748   \\
		& B & 0.956298 & 0.959205 & 0.930419 & -0.001179 & -0.000550 & -0.002195 & -0.001821  & -0.000839 & 0.003026    \\
		\midrule
		\multirow{3}{*}{Lake}    & R & 0.954300 & 0.950933 & 0.938652 & -0.002712 & -0.002231 & -0.007473 & 0.008294   & 0.002791  & 0.005007    \\
		& G & 0.968942 & 0.972935 & 0.957351 & -0.006042 & 0.005319  & -0.003452 & -0.005502  & -0.003010 & -0.000647   \\
		& B & 0.968070 & 0.969546 & 0.951428 & 0.005635  & -0.003997 & 0.002186  & 0.010300   & -0.003263 & 0.001231    \\
		\midrule
		\multirow{3}{*}{Peppers}  & R & 0.968357 & 0.956775 & 0.950391 & -0.000807 & -0.004586 & -0.001847 & -0.002638  & 0.007745  & 0.005794    \\
		& G & 0.986690 & 0.972400 & 0.964351 & 0.004994  & 0.011397  & 0.008190  & 0.003480   & 0.004427  & 0.004097    \\
		& B & 0.972810 & 0.959227 & 0.947239 & -0.003806 & 0.007530  & -0.006979 & -0.003785  & -0.004886 & -0.001596   \\
		\midrule
		\multirow{3}{*}{Mandrill} & R & 0.853960 & 0.910030 & 0.841495 & 0.000305  & -0.009033 & 0.004641  & 0.010308   & 0.004073  & 0.007677    \\
		& G & 0.771212 & 0.858922 & 0.739454 & 0.000354  & 0.006221  & -0.000766 & -0.008846  & -0.006407 & 0.000659    \\
		& B & 0.864290 & 0.892019 & 0.825914 & 0.005260  & -0.007633 & 0.001064  & 0.002351   & -0.010925 & 0.009007    \\
		\midrule
		\multirow{3}{*}{Splash}   & R & 0.985901 & 0.992967 & 0.980009 & 0.012251  & -0.003002 & -0.005989 & -0.002299  & 0.005270  & 0.005558    \\
		& G & 0.979680 & 0.981868 & 0.965632 & 0.006494  & 0.003150  & -0.006994 & 0.001839   & -0.002984 & 0.001348    \\
		& B & 0.959442 & 0.983775 & 0.948882 & 0.015445  & 0.003662  & -0.009827 & -0.006020  & -0.010502 & 0.002206    \\
		\midrule
		\multicolumn{11}{l}{H: horizontal correlation coefficient; V: vertical correlation coefficient; D: diagonal correlation coefficient.}\\
		\bottomrule
		\label{Tab:statisticalEvaluation}
	\end{tabular}
\end{table*}

\subsection{Correlation Evaluation}
Since two adjacent pixels in an image usually have high correlation, the cryptosystems should decrease the correlation to resist statistical attacks \cite{ye2012efficient}. Therefore, the correlation between two adjacent pixels in plain and cipher images are analyzed. We first randomly select 6000 pairs of two horizontally, vertically, and diagonally adjacent pixels from the plain and cipher images. The correlation distribution of red, green, and blue channels in plain image shown in Fig. \ref{Fig:correlationEvaluation} (a), (b), (c), respectively. The correlation distribution of red, green, and blue channels in cipher image are plotted in Fig. \ref{Fig:correlationEvaluation} (d), (e), (f), respectively. The correlation distribution of red, green, and blue channels in cipher image shown in Fig. \ref{Fig:uniformityEvaluation} (i) are drawn in Fig. \ref{Fig:correlationEvaluation} (g), (h), (i), respectively.

Then, the correlation coefficient between adjacent pixels are measured.  The correlation coefficient $r_{x,y}$ of adjacent pixels can be calculated out according to the following equation \cite{rhouma2009ocml}:
\begin{equation}
	r_{x,y}=\frac{\mathrm{cov}(x,y)}{\sqrt{D(x)D(y)}},
\end{equation}
where $x$ and $y$ are the values of two adjacent pixels in the image, $\mathrm{cov}(x,y)$ is given by
\begin{equation}
	\mathrm{cov}(x,y)=\frac{1}{N}\sum_{i=1}^N(x_i-E(x))(y_i-E(y)),
\end{equation}
$N$ is the total number of pixels selected from the image for the calculation, $E(x)$ and $D(x)$ are defined as follows:
\begin{equation}
	E(x) = \frac{1}{N}\sum_{i=1}^Nx_i,
\end{equation}
\begin{equation}
	D(x) = \frac{1}{N}\sum_{i=1}^N(x_i-E(x))^2.
\end{equation}
When the correlation coefficient of a cipher image is lower, the resistance capability of the encryption algorithm to statistical analysis attacks is stronger.
To assess the correlation property of the proposed strategy,
we encrypt a set of plain images using PLCM and 2DLASM, randomly select 20000 adjacent pixels in horizontal, vertical, and diagonal directions from the plain and cipher images, calculate the correlation coefficients, and list the results in Tab. \ref{Tab:statisticalEvaluation}. The deployed cryptosystesm, clearly, significantly decrease the correlation of the plain images.

\subsection{Randomness Evaluation}
In this section, three widely used methods, i.e., information entropy, $(k, T_B)$ local Shannon entropy, and NIST statistical tests, are employed to evaluate the randomness properties of the proposed strategy.
The information entropy is defined to quantify the information randomness \cite{belazi2017efficient}. For a given information $m$, the entropy $H(m)$ can be calculated as follows:
\begin{equation}
	H(m) = \sum_{i=1}^{N}p(m_i)\mathrm{log}\frac{1}{p(m_i)},
\end{equation}

\begin{table}[!thpb]\small
\caption{Information entropy $H$ and $(k,T_B)$-local Shannon entropy $\overline{H}_{k,T_B}$ of plain and cipher images.}
\begin{tabular}{cl|ccc}
\toprule
\multirow{2}{0.6cm}{Test} 						& \multirow{2}{2.6cm}{Image}  	   & \multicolumn{3}{c}{Channel}\\
\cmidrule(r){3-5}
												&                     & R          & G          & B\\
\midrule
\multirow{18}{0.6cm}{H}         				& Lena                & 7.253102   & 7.594038   & 6.968427\\
												& Lena (PLCM)         & 7.999369   & 7.999326   & 7.999449\\
												& Lena (2DLASM)       & 7.999259   & 7.999282   & 7.999269\\
\cmidrule(r){2-5}
												& Airplane            & 6.717765   & 6.798979   & 6.213774\\
												& Airplane (PLCM)     & 7.999219   & 7.999286   & 7.999361\\
												& Airplane (2DLASM)   & 7.999299   & 7.999234   & 7.999291\\
\cmidrule(r){2-5}
												& Lake                & 7.312388   & 7.642854   & 7.213642\\
												& Lake (PLCM)         & 7.999353   & 7.999302   & 7.999199\\
												& Lake (2DLASM)       & 7.999271   & 7.999373   & 7.999268\\
\cmidrule(r){2-5}
												& Peppers             & 7.338827   & 7.496253   & 7.058306\\
												& Peppers (PLCM)      & 7.999328   & 7.999317   & 7.999370\\
												& Peppers (2DLASM)    & 7.999292   & 7.999353   & 7.999245\\
\cmidrule(r){2-5}
												& Mandrill            & 7.706672   & 7.474432   & 7.752217\\
												& Mandrill (PLCM)     & 7.999302   & 7.999302   & 7.999221\\
												& Mandrill (2DLASM)   & 7.999201   & 7.999319   & 7.999403\\
\cmidrule(r){2-5}
												& Splash              & 6.948061   & 6.884455   & 6.126452\\
												& Splash (PLCM)       & 7.999305   & 7.999238   & 7.999263\\
												& Splash (2DLASM)     & 7.999346   & 7.999397   & 7.999225\\
\midrule
\multirow{18}{0.6cm}{$\overline{H}_{k,T_B}$}    & Lena                & 5.911418   & 6.378169   & 6.044504\\
												& Lena (PLCM)         & 7.901188   & 7.900454   & 7.902854\\
												& Lena (2DLASM)       & 7.903580   & 7.902076   & 7.901568\\
\cmidrule(r){2-5}
												& Airplane            & 5.371063   & 5.348903   & 4.918050\\
												& Airplane (PLCM)     & 7.903643   & 7.901751   & 7.902929\\
												& Airplane (2DLASM)   & 7.903149   & 7.904907   & 7.901914\\
\cmidrule(r){2-5}
												& Lake                & 6.129053   & 6.348979   & 5.847770\\
												& Lake (PLCM)         & 7.906276   & 7.901353   & 7.901770\\
												& Lake (2DLASM)       & 7.902562   & 7.900261   & 7.901848\\
\cmidrule(r){2-5}
												& Peppers             & 5.975824   & 6.057454   & 5.870517\\
												& Peppers (PLCM)      & 7.903389   & 7.905916   & 7.901726\\
												& Peppers (2DLASM)    & 7.902408   & 7.899511   & 7.904691\\
\cmidrule(r){2-5}
												& Mandrill            & 6.578668   & 6.688874   & 6.802597\\
												& Mandrill (PLCM)     & 7.903176   & 7.901764   & 7.903579\\
												& Mandrill (2DLASM)   & 7.901823   & 7.901584   & 7.903498\\
\cmidrule(r){2-5}
												& Splash              & 4.150002   & 5.184026   & 4.560897\\
												& Splash (PLCM)       & 7.903036   & 7.902193   & 7.903928\\
												& Spalsh (2DLASM)     & 7.900637   & 7.904737   & 7.902950\\
\bottomrule
\label{Tab:informationEntropy}
\end{tabular}
\end{table}

\newpage
\noindent where $N$ is the total number of symbols, and $p(m_i)$ is the occurrence probability of symbol $m_i$. For a random figure $F$, the entropy $H(F)$ equals to $8$. A cryptosystem, therefore, should generate cipher image with information entropy close to $8$.

The $(k, T_B)$ local Shannon entropy is utilized to measure the local randomness of images \cite{wu2013local}, it is defined as follows:
\begin{equation}
	\overline{H}_{k,T_B}=\sum_{i=1}^k\frac{H(M_i)}{k},
\end{equation}
where $M_i~(i\in\{1,2,...,k\})$ are the randomly selected non-overlapping image sub-blocks, each block has $T_B$ pixels, $H(M_i)$ is the information entropy of block $M_i$.
We encrypt a set of plain images with randomly selected key $K$ using PLCM and 2DLASM, respectively, list the information entropy results in Tab. \ref{Tab:informationEntropy}. Then we set $k$ and $T_B$ to $30$ and $1936$ according to Refs. \cite{artiles2019image, xian2021fractal}, randomly select non-overlapping blocks from the plain and the generated cipher images, calculate $\overline{H}_{k,T_B}$. and list the results in Tab. \ref{Tab:informationEntropy}.

NIST statistical test suite (version 800-22) can be used to test if a sequence generated by random or pseudorandom number generators is suitable for cryptographic applications \cite{rukhin2001statistical}. It consists fifteen statistical tests focusing on distinct types of nonrandomness that may exist in a bit sequence. The tests are used to determine the acceptance of the hypothesis of ideal randomness with significance level $\alpha$, which is set to $0.01$. For each test, $P$-value and pass rate $r_p$ are regarded as the criteria results. When $P$-value is larger than the significance level $\alpha$, the sequence is regard as random. In video encryption, the assistant threads take a subframe, generate byte sequence, and perform diffusion operation on the subframe. According to the encryption process, we use $8$ assistant threads $T_a^i~i\in\{1,2,...,8\}$ to generate enough byte sequences $B_1, B_2,...,B_8$ using PLCM and 2DLASM, respectively. And perform NIST tests on the generated bytes. The results are shown in Tab. \ref{Tab:NISTTests}.

\begin{table}[!thpb]\small
\caption{Results of NIST tests.}
\begin{tabular}{p{3.7cm}p{1cm}p{0.6cm}p{1cm}p{0.6cm}}
\toprule
\multirow{2}{3.7cm}{Statistical Test}  	& \multicolumn{2}{c}{PLCM}       	&  \multicolumn{2}{c}{2DLASM}\\
\cmidrule(r){2-5}
										& $P$-value         & $r_p$(\%)     & $P$-value     & $r_p$(\%)\\
\midrule
Frequency                       		& 0.181557          & 99            & 0.657933      & 98\\
Block Frequency                	 		& 0.474986          & 99            & 0.955835      & 99\\
Cumulative Sums$^*$             		& 0.452541          & 99            & 0.717629      & 98\\
Runs                            		& 0.350485          & 99            & 0.115387      & 97\\
Longest Run                     		& 0.883171          & 97            & 0.171867      & 100\\
Rank                            		& 0.366918          & 99            & 0.554420      & 100\\
FFT                             		& 0.897763          & 100           & 0.595549      & 100\\
Non Overlapping Template$^*$    		& 0.486724          & 99            & 0.510463      & 98\\
Overlapping Template            		& 0.366918          & 98            & 0.883171      & 96\\
Universal                       		& 0.455937          & 100           & 0.574903      & 97\\
Approximate Entropy             		& 0.983453          & 99            & 0.249284      & 98\\
Random Excursions$^*$           		& 0.415715          & 99            & 0.258140      & 99\\
Random Excursions Variant$^*$   		& 0.442424          & 99            & 0.276585      & 99\\
Serial$^*$                      		& 0.428515          & 99            & 0.944186      & 99\\
Linear Complexity               		& 0.137282          & 96            & 0.058984      & 100\\
\midrule
\multicolumn{5}{l}{Significance level $\alpha=0.01$; $r_p$: pass rate; $*$: mean value.}\\
\bottomrule
\label{Tab:NISTTests}
\end{tabular}
\end{table}

\begin{figure*}[b]
	\centering
	\includegraphics[width=0.98\textwidth]{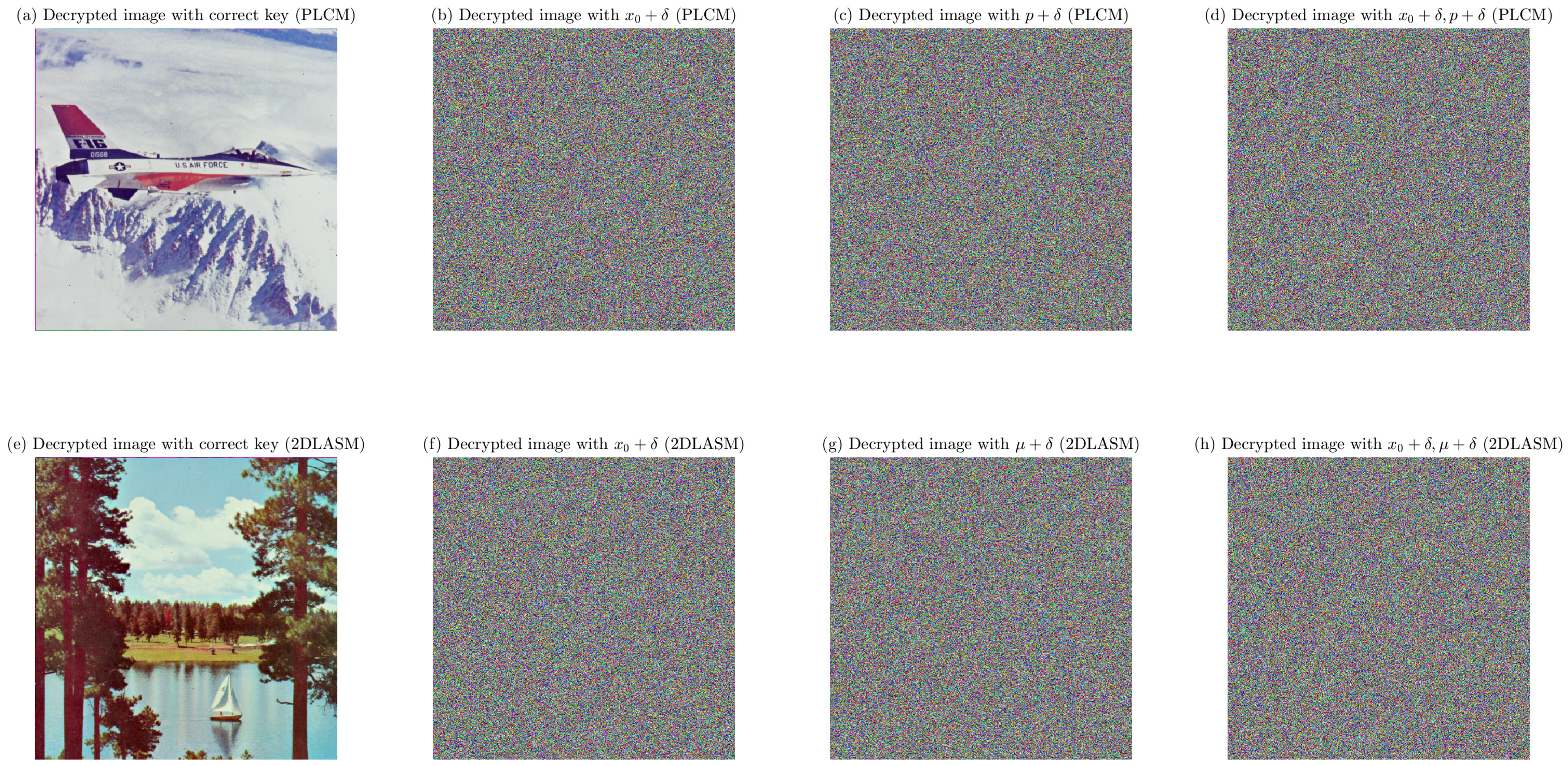}
	\caption{Key sensitivity analysis. (a) decrypt image using PLCM with correct key, (b) decrypt image using PLCM with $x_0+\delta$, (c) decrypt image using PLCM with $p+\delta$, (d) decrypt image using PLCM with $x_0+\delta$ and $p+\delta$, (e) decrypt image using 2DLASM with correct key, (f) decrypt image using 2DLASM with $x_0+\delta$, (g) decrypt image using 2DLASM with $\mu+\delta$, (h) decrypt image using 2DLASM with $x_0+\delta$ and $p+\delta$. The increment $\delta$ is set to $1\times10^{-9}$.}
	\label{Fig:keySensitivity}
\end{figure*}

\section{Security Analysis}
\subsection{Key Space And Sensitivity}

One of the most important issue of the security of cryptosystems is the key. According to the Kerckhoff's principle, a cryptosystem should be secure even though everything about the system, except the key, is known by the cracker \cite{wang2012novel}. If the key is poorly chosen or the key space is too small, no matter how strong the encryption algorithm might be, the cryptosystem can be easily cracked. It is generally accepted that the key space of a cryptosystem should be large that $2^{100}$ to provide a sufficient security against ciphertext only attaks \cite{alvarez2006some}. For the proposed strategy, clearly, the key is the parameters that used to initialize $\mathrm{PRBG}_m$ of the main thread $T_m$. In PLCM based cryptosystesm, four double precision floating-point numbers are taken as the initial conditions and control parameters to initialize two PLCMs. In the 2DLASM based cryptosystem, three double precision floating-point numbers are input as the initial condition $x_0$, control parameter $\mu$, and $y_0$ to initialize 2DLASM. The key space of PLCM and 2DLASM based cryptosystems, therefore, are $2^{256}$ and $2^{192}$, respectively. This clearly, provides the resistance to ciphertext only attacks. The users can employ more complex $\mathrm{PRBG}$ to obtain larger key space, resulting in higher security.

In addition, the cryptosystem should have high sensitivity to the change of key in encryption and decryption process. That is, even one bit of the key is changed, the attacker cannot obtain any information from the decrypted image. To analyze the key sensitivity of the proposed strategy, we randomly select key, use the deployed cryptosystems to ecnrypt an iamge, followed by decrypting the cipher image with the correct key. The images decrypted with PLCM and 2DLASM are plotted in Fig. \ref{Fig:keySensitivity} (a) and (e), respectively. Then we slight change the key by adding an increment $\delta=0.000000001$ on initial condition and control parameter, decrypted the cipher image with the slightly modified Key. The images decrypted by PLCM based cryptosystem using $x_0+\delta$, $p+\delta$, and $x_0+\delta, p+\delta$ are shown in Fig. \ref{Fig:keySensitivity} (b), (c), and (d), respectively. The images decrypted by 2DLASM based cryptosystem using $x_0+\delta$, $\mu+\delta$, and $x_0+\delta, \mu+\delta$ are shown in Fig. \ref{Fig:keySensitivity} (f), (g), and (h), respectively. We also calculate the correlation coefficients between the images decrypted with the correct key and the images decrypted with the slightly changed key, and list the results in Tab. \ref{Tab:keySensitivity}.

\begin{table}[!thpb]\small
\caption{Correlation coefficients between decrypted images plotted in Fig. \ref{Fig:keySensitivity}.}
\begin{tabular}{ll|ccc}
\toprule
\multirow{2}{1.1cm}{PRBG} 	& \multirow{2}{1.6cm}{Images}  					 & \multicolumn{3}{c}{Channel}\\
\cmidrule(r){3-5}
							&                                                & R          & G         & B\\
\midrule
\multirow{6}{1.0cm}{PLCM}   & Fig. \ref{Fig:keySensitivity} (a), (b)         & -0.002588  & 0.005312   & 0.002903\\
							& Fig. \ref{Fig:keySensitivity} (a), (c)         & -0.001788  & 0.001664   & -0.001074\\
							& Fig. \ref{Fig:keySensitivity} (a), (d)         & -0.002994  & 0.001132   & 0.002672\\
							& Fig. \ref{Fig:keySensitivity} (b), (c)         & 0.004292   & -0.001097  & -0.000042\\
							& Fig. \ref{Fig:keySensitivity} (b), (d)         & 0.002267   & 0.001098   & 0.002168\\
							& Fig. \ref{Fig:keySensitivity} (c), (d)         & 0.000715   & 0.004356   & 0.000954\\
\midrule
\multirow{6}{1.0cm}{2DLASM} & Fig. \ref{Fig:keySensitivity} (e), (f)         & -0.002142  & 0.000862   & 0.000712\\
							& Fig. \ref{Fig:keySensitivity} (e), (g)         & -0.002620  & 0.001053   & -0.000951\\
							& Fig. \ref{Fig:keySensitivity} (e), (h)         & 0.003072   & -0.000513  & 0.001539\\
							& Fig. \ref{Fig:keySensitivity} (f), (g)         & -0.001342  & -0.000709  & 0.002444\\
							& Fig. \ref{Fig:keySensitivity} (f), (h)         & 0.004969   & -0.001565  & 0.000370\\
							& Fig. \ref{Fig:keySensitivity} (g), (h)         & 0.003579   & 0.001962   & 0.000740\\
\bottomrule
\label{Tab:keySensitivity}
\end{tabular}
\end{table}

\begin{table*}[b]\small
\caption{Results of NPCR and UACI}
\begin{tabular}{lll|ccc|ccc|ccc}
\toprule
& & & \multicolumn{9}{c}{Channel}\\
\cmidrule(r){4-12}
Test                       	& Image                             & PRBG     	& \multicolumn{3}{c|}{R} 		    & \multicolumn{3}{c|}{G} 	        & \multicolumn{3}{c}{B}\\
\cmidrule(r){4-12}
							& 						   			&			& Min      	& Max     	& Average 	& Min     	& Max     	& Average 	& Min     	& Max     	& Average \\
\midrule
\multirow{12}{1cm}{NPCR}   	& \multirow{2}{1cm}{Lena}           & CPLCM    	& 99.5770	& 99.6334 	& 99.6093 	& 99.5762 	& 99.6361 	& 99.6096 	& 99.5781 	& 99.6372 	& 99.6048\\
							&                                   & 2DLASM   	& 98.9208 	& 99.6376 	& 99.5696 	& 99.1913 	& 99.6395 	& 99.5667 	& 99.2020 	& 99.6414 	& 99.5839\\
\cmidrule(r){2-12}
							& \multirow{2}{1cm}{Airplane}      	& CPLCM    	& 99.5800	& 99.6471 	& 99.6092 	& 99.5789 	& 99.6426 	& 99.6093 	& 99.5861 	& 99.6418 	& 99.6091\\
							&                                   & 2DLASM   	& 98.4215 	& 99.6349 	& 99.5496 	& 98.4520 	& 99.6433 	& 99.5409 	& 98.4180 	& 99.6403 	& 99.5138\\
\cmidrule(r){2-12}
							& \multirow{2}{1cm}{Lake}           & CPLCM    	& 98.5302 	& 99.6841 	& 99.5808 	& 96.8250 	& 99.6315 	& 99.5587 	& 98.4116 	& 99.6395 	& 99.5757\\
							&                                   & 2DLASM   	& 98.6176 	& 99.6284 	& 99.5616 	& 99.1982 	& 99.6384 	& 99.5949 	& 99.1920 	& 99.6445 	& 99.5709\\
\cmidrule(r){2-12}
							& \multirow{2}{1cm}{Peppers}        & CPLCM    	& 98.4627 	& 99.6361 	& 99.5762 	& 98.4226 	& 99.6510 	& 99.5301 	& 99.2138 	& 99.6407 	& 99.5908\\
							&                                   & 2DLASM   	& 97.2801 	& 99.6559 	& 99.5536 	& 97.3145 	& 99.6437 	& 99.5582 	& 98.6259 	& 99.6391 	& 99.5693\\
\cmidrule(r){2-12}
							& \multirow{2}{1cm}{Mandrill}       & CPLCM    	& 97.2404 	& 99.6464 	& 99.5532 	& 98.8369 	& 99.6407 	& 99.5809 	& 98.4257 	& 99.6304 	& 99.5529\\
							&                                   & 2DLASM   	& 98.4825 	& 99.7314 	& 99.5763 	& 98.7518 	& 99.6433 	& 99.5658 	& 98.6263 	& 99.6418 	& 99.5753\\
\cmidrule(r){2-12}
							& \multirow{2}{1cm}{Splash}         & CPLCM    	& 98.6454 	& 99.6368 	& 99.5792 	& 98.4016 	& 99.6349 	& 99.5604 	& 98.8396 	& 99.6330 	& 99.5838\\
							&                                   & 2DLASM   	& 98.4062 	& 99.6349 	& 99.5779 	& 98.4413 	& 99.6376 	& 99.5171 	& 98.4142 	& 99.6357 	& 99.5463\\		
\midrule

\multirow{12}{1cm}{UACI}    & \multirow{2}{1cm}{Lena}           & CPLCM    & 33.3000 	& 33.5896 	& 33.4585 	& 33.3756 	& 33.6102 	& 33.4633 	& 33.3486 	& 33.5817 	& 33.4616\\
							&                                   & 2DLASM   & 33.3106 	& 33.6019 	& 33.4648 	& 33.3245 	& 33.5847 	& 33.4678 	& 33.3580 	& 33.5815 	& 33.4674\\
\cmidrule(r){2-12}
							& \multirow{2}{1cm}{Airplane}       & CPLCM    & 33.3524 	& 33.5590 	& 33.4615 	& 33.3716 	& 33.5565 	& 33.4650 	& 33.3379 	& 33.5907 	& 33.4682\\
							&                                   & 2DLASM   & 33.3349 	& 33.5936 	& 33.4680 	& 33.3895 	& 33.6084 	& 33.4678 	& 33.3561 	& 33.5746 	& 33.4734\\
\cmidrule(r){2-12}
							& \multirow{2}{1cm}{Lake}           & CPLCM    & 33.3271 	& 33.5673 	& 33.4630 	& 33.3637 	& 33.5900 	& 33.4675 	& 33.3601 	& 33.5879 	& 33.4668\\
							&                                   & 2DLASM   & 33.3629 	& 33.5763 	& 33.4700 	& 33.3610 	& 33.5877 	& 33.4558 	& 33.3493 	& 33.5729 	& 33.4652\\
\cmidrule(r){2-12}
							& \multirow{2}{1cm}{Peppers}        & CPLCM    & 33.3690 	& 33.5836 	& 33.4623 	& 33.3517 	& 33.5588 	& 33.4577 	& 33.3644 	& 33.5995 	& 33.4632\\
							&                                   & 2DLASM   & 33.3299 	& 33.5721 	& 33.4585 	& 33.3322 	& 33.5540 	& 33.4664 	& 33.3713 	& 33.5711 	& 33.4687\\
\cmidrule(r){2-12}
							& \multirow{2}{1cm}{Mandrill}       & CPLCM    & 33.3343 	& 33.5772 	& 33.4643 	& 33.3393 	& 33.5597 	& 33.4644 	& 33.3561 	& 33.5829 	& 33.4603\\
							&                                   & 2DLASM   & 33.3706 	& 33.5938 	& 33.4720 	& 33.3503 	& 33.5429 	& 33.4555 	& 33.3110 	& 33.6242 	& 33.4626\\
\cmidrule(r){2-12}
							& \multirow{2}{1cm}{Splash}         & CPLCM    & 33.3785 	& 33.5953 	& 33.4662 	& 33.3411 	& 33.5438 	& 33.4536 	& 33.3535 	& 33.5860 	& 33.4630\\
							&                                   & 2DLASM   & 33.3427 	& 33.5976 	& 33.4647 	& 33.3602 	& 33.6073 	& 33.4600 	& 33.3805 	& 33.5635 	& 33.4619\\
\bottomrule
\label{Tab:differentialAttack}
\end{tabular}
\end{table*}

\subsection{Resistance To Differential Attacks}
To resist differential attacks, the encryption algorithms should have high sensitivity to the plain image, that is, a minor change in the plain image will lead to a completely change in the cipher image \cite{khan2019novel}. To evaluate the ability of the proposed strategy to resist such attacks, Number of Pixels Change Rate (NPCR) and Unified Average Changing Intensity (UACI) are calculated \cite{luo2019novel}. NPCR can be calculated according to the following equation:
\begin{equation}
	\label{Equ:NPCR}
	\mathrm{NPCR}=\sum\limits_{i=0}^w\sum\limits_{j=0}^h\frac{D(i, j)}{w\times h}\times 100\%,
\end{equation}
where $w$ and $h$ denote the width and the height of the images, $D(i,j)$ is defined as follows:
\begin{equation}
	D(i,j)=
	\left\{
	\begin{array}{lr}
		1, ~F_c^1[i,j] \neq F_c^2[i,j],\\
		0, ~F_c^1[i,j] = F_c^2[i,j],
	\end{array}
	\right.
\end{equation}
$F_c^1$ and $F_c^2$ are two cipher images generated with the plain image and the slightly changed plain image, respectively, $F_c[i,j]$ is the pixel value. UACI is given by:

\newpage
\begin{equation}
	\label{Equ:UACI}
	\mathrm{UACI}=\sum_{i=0}^w\sum_{j=0}^h\frac{|F_c^1[i,j] - F_c^2[i,j]|}{w\times h\times255}\times 100\%,
\end{equation}
According to Refs. \cite{mansouri2020novel, talhaoui2021new}, for an image of size $512\times512$, when the significance level is equal to 0.05, the expected values of NPCR and UACI are $99.5893\%$ and $[33.3730\%, 33.5541\%]$, respectively. An image encryption scheme passes the diffusion property test when NPCR is higher than the expected value, and UACI falls within the interval.

In the proposed strategy, despite all assistant threads perform diffusion operations independently, they take the last pixel of the next subframe as the diffusion seed to reconstruct the relationship between all subframes. This guarantees that one pixel changed in any subframe, will result in a completely different cipher frame.
To fully evaluate the resistance of the proposed strategy to differential attacks, we generate byte sequences, encrypt a set of plain images, randomly select and change a pixel in the plain images, encrypt the modified images with the same byte sequences, calculate NPCR and UACI between the generated cipher images.
For each plain image, we repeat above steps for 100 times, fetch the minimum, maximum, and average values, and list the results in Tab. \ref{Tab:differentialAttack}.

\begin{figure*}[t]
	\centering
	\includegraphics[width=0.98\textwidth]{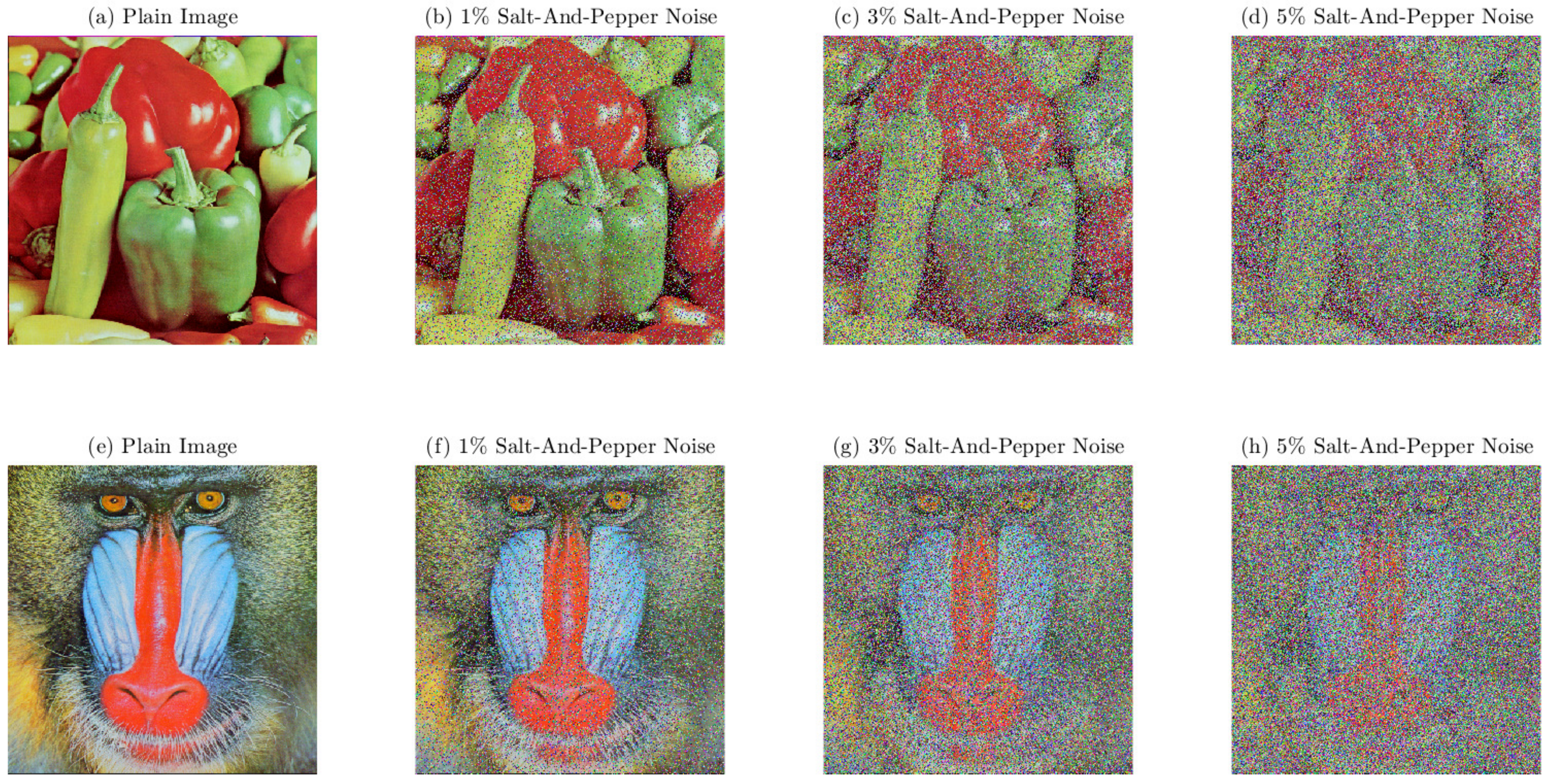}
	\caption{Resistance to noise. (a) plain image peppers, (b)-(d) decrypt the cipher image with $1\%$, $3\%$, $5\%$ salt-and-pepper noise using PLCM, (e) plain image mandrill, (f)-(h) decrypt the cipher image with $1\%$, $3\%$, $5\%$ salt-and-pepper noise using 2DLASM,.}
	\label{Fig:saltAndPepperNoise}
\end{figure*}

\begin{figure*}[!thbp]
	\centering
	\includegraphics[width=0.98\textwidth]{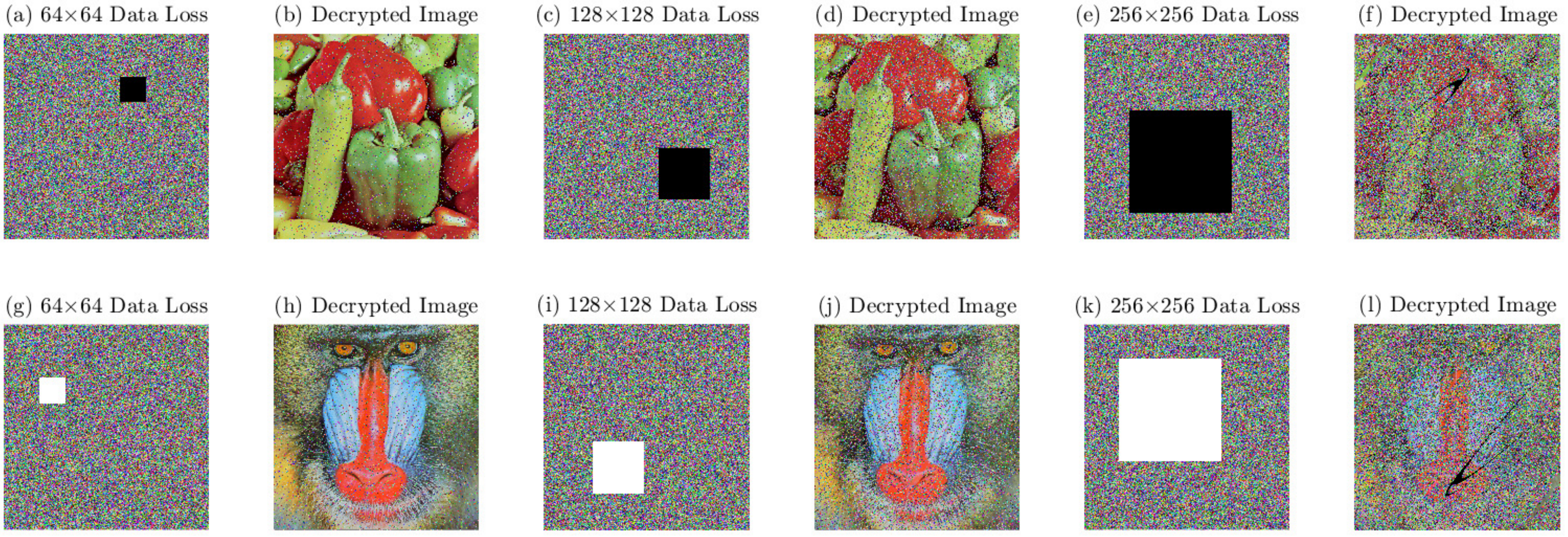}
	\caption{Resistance to data loss. (a)-(f) cipher images with different data losses and the corresponding images decrypted with PLCM, (g)-(j) cipher images with different data losses and the corresponding images decrypted with 2DLASM.}
	\label{Fig:dataLoss}
\end{figure*}

\subsection{Resistance To Noise And Data Loss}

Images or video frames may be affected by noise while transmitting over the channel. When a cipher image has noise or losses part of data, the cyrptosystem needs to recover the original image with high visual quality \cite{chen2022fast}. To evaluate the robustness, we use the two deployed cryptosystems to encrypt a plain image with randomly selected keys, add $1\%$, $3\%$, and $5\%$ salt-and-pepper noise \cite{azzeh2018salt} to the generated cipher images, use the same keys to decrypt the cipher images, and plot the plain and decrypted images in Fig. \ref{Fig:saltAndPepperNoise}. Although the decrypted image are changed when there exists noise in the cipher image, the approximate information of the plain image is preserved, even when the slat-and-pepper noise as high as $5\%$.

In real application, image cropping is very common, which may lead to data loss \cite{gan2019chaotic}. To assess the capability of the proposed strategy to resist the data loss, similarly, we use the deployed cryptosystems to encrypt a plain image, randomly select and remove $64\times 64$, $128\times 128$, or $512\times 512$ blocks from the cipher images, replace the removed block with white or black blocks, decrypt the modified cipher images with the same keys, and show the results in Fig. \ref{Fig:dataLoss}. Clearly, the contours of the plain images are preserved, even though $256\times 256$ blocks are removed from the cipher images. That is, when there exist $25\%$ data loss in the cipher image, the deployed cryptosystems can still recover the information of the plain image.
The proposed strategy, therefore, can resist noise and data loss.

\begin{figure*}[t]
	\centering
	\includegraphics[width=0.96\textwidth]{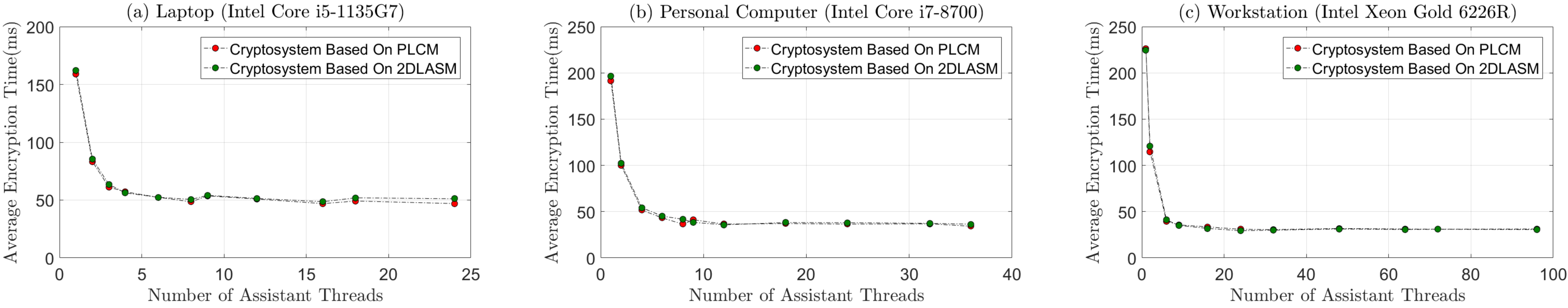}
	\caption{Relationship between the number of assistant threads and the encryption speed, (a) results of the laptop (Intel Core i5-1135G7), (b) results of the personal computer (Intel Core i7-8700G7), (c) results of the workstation (Intel Xeon Gold 6226R).}
	\label{Fig:SpeedEvaluation}
\end{figure*}

\begin{figure*}[!thbp]
	\centering
	\includegraphics[width=0.96\textwidth]{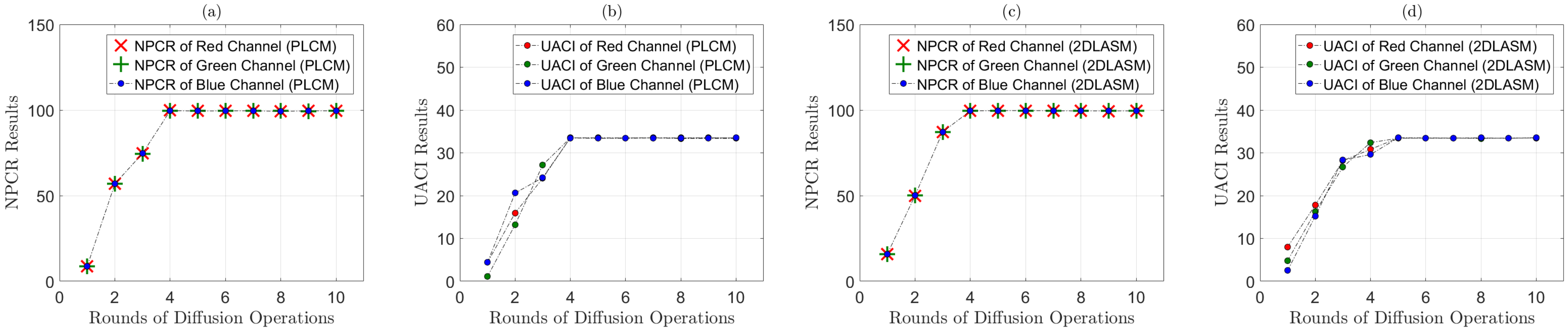}
	\caption{Relationship between the rounds of diffusion operations and NPCR, UACI results, (a) NPCR results of the cryptosystem based on PLCM, (b) UACI results of the cyrptosystem based on PLCM, (c) NPCR results of the cryptosystem based on 2DLASM, (d) UACI results of the cyrptosystem based on 2DLASM.}
	\label{Fig:DiffusionAnalysis}
\end{figure*}

\begin{figure*}[!thbp]
	\centering
	\includegraphics[width=0.98\textwidth]{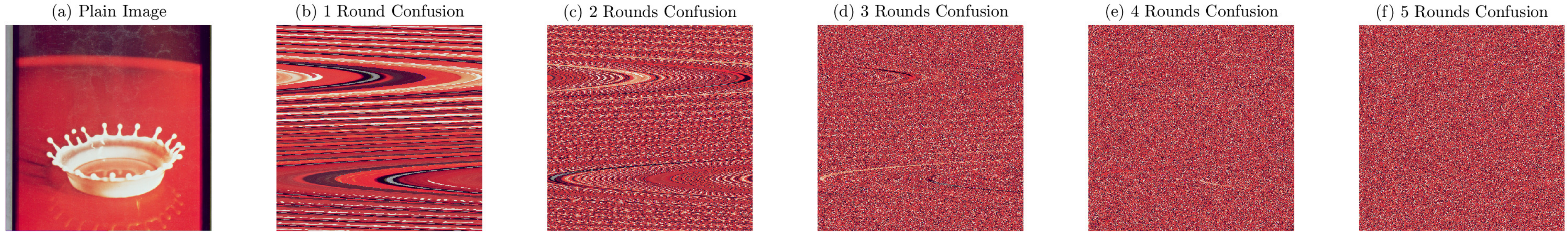}
	\caption{Results of confusion operations. (a) plain image splash, (b)-(f) images after different rounds of confusion operations, .}
	\label{Fig:confusionResults}
\end{figure*}

\section{Parameter Setup}

According to the encryption process, clearly, the number of assistant threads $n$ and the rounds of confusion and diffusion $r$ significantly affect the  the encryption speed and the statistical performance of the deployed cryptosystems. In all experiments carried out in this paper, $n$ is set to 8, 12, and 32 for laptop, personal computer, and workstation, respectively, and $r$ is set to 5. The reasons for such settings are analyzed in this section.

\begin{figure}[!thbp]
	\centering
	\includegraphics[width=0.48\textwidth]{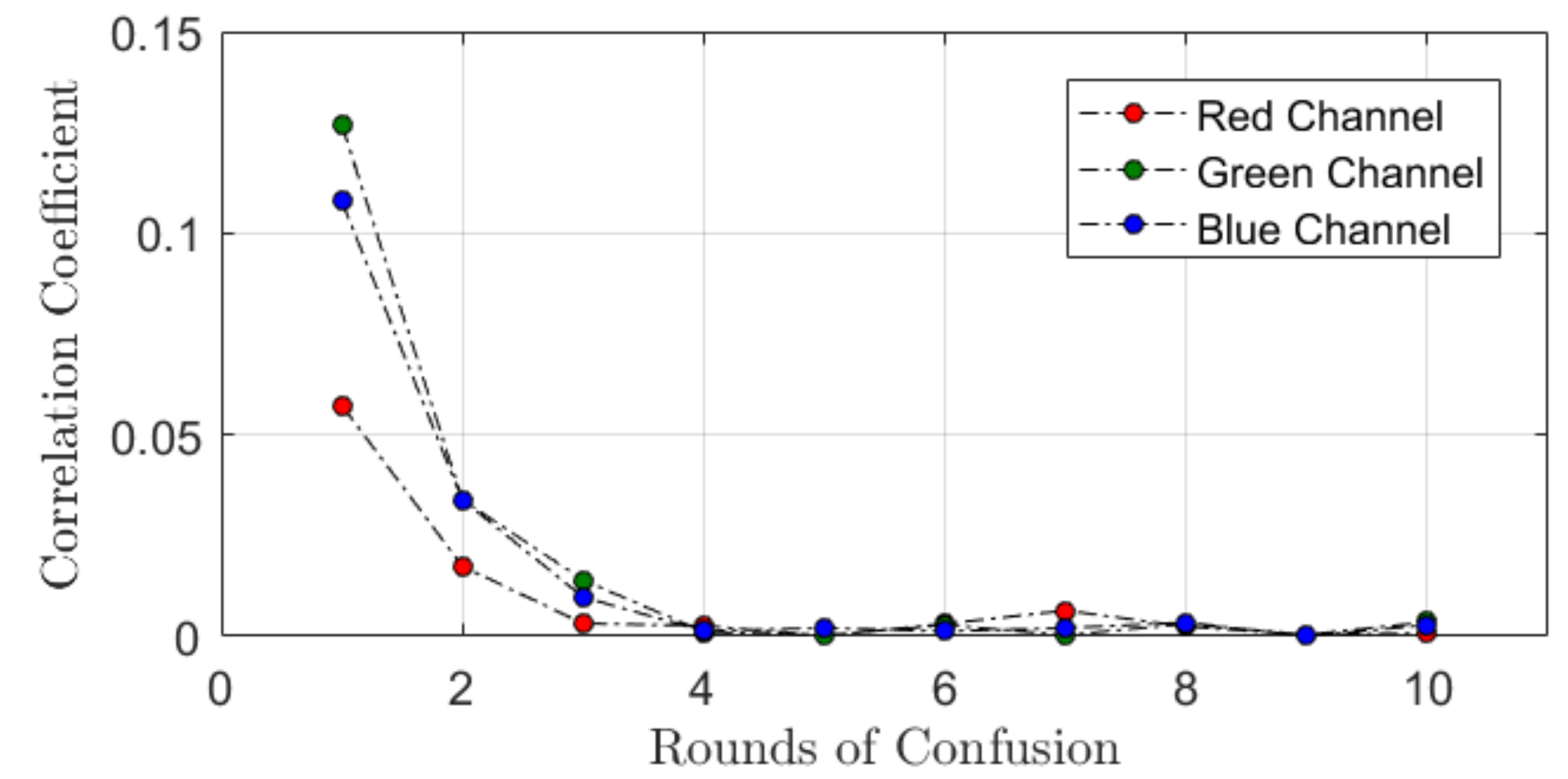}
	\caption{Correlation coefficient between the plain image and the images after different rounds of confusion operations.}
	\label{Fig:confusionAnalysis}
\end{figure}

We use different number of assistant threads to encrypt a plain video of size $576\times 576$ using three hardware platforms. The relationship between the average encryption time and the number of assistant threads are drawn in Fig. \ref{Fig:SpeedEvaluation}. 
Set $n$ to the maximum number of threads supported by CPU is unnecessary, however, can guarantee the deployed cryptosystems reach their fastest encryption speed.

The purposes of diffusion and confusion operations are to improve the plaintext sensitivity of the cryptosystems and decrease the correlation of the plain images. 
We, thus, generate byte sequences, perform 10 rounds of diffusion operations on a plain image, randomly select a pixel, change its value, perform 10 rounds of diffusion operations on the modified image with the same byte sequences, and calculate NPRC and UACI between the generated cipher images. The results are plotted in Fig. \ref{Fig:DiffusionAnalysis}.
Similarly, we perform 10 rounds of confusion operations on a plain image. The plain image and the scrambled images after 1-5 rounds of confusion operations are shown in Fig.  \ref{Fig:confusionResults}. The correlation coefficients between the plain image and the scrambled images after different rounds of confusion operations are drawn in Fig. \ref{Fig:confusionAnalysis}. When $r$ is equal to $5$, clearly, NPCR, UACI, and the correlation coefficient reach the upper and lower bounds, respectively.

\begin{table*}[b]\small
\caption{Speed comparison to previous works}
\begin{tabular}{p{1.4cm}p{0.4cm}p{0.4cm}p{4.6cm}p{1.2cm}p{1.6cm}p{1cm}p{2.8cm}p{0.6cm}}
\toprule
Algorithm                                       &$r_c$                  &$r_d$                  &CPU                                                &Memory                    &Video Size                             &CS                             &AET(ms)            &RTE\\
\midrule
\multirow{3}{1.6cm}{Ref. \cite{valli2017chaos}} &\multirow{3}{0.6cm}{1} &\multirow{3}{0.6cm}{1} &\multirow{3}{4.6cm}{Pentium 4@3.2GHz}              &\multirow{3}{1.2cm}{4GB}  &\multirow{3}{1.6cm}{$240\times360$}    &\multirow{3}{1cm}{RGB}         &1059 (8D Map)      &$\times$\\
												&                       &                       &                                                   &                          &                                       &                               &2314 (12D Map)     &$\times$\\
												&                       &                       &                                                   &                          &                                       &                               &1357 (Ikeda DDE)   &$\times$\\
\midrule
Ref. \cite{ganeshkumar2019new}                  &1                      &1                      &--                                                 &--                        &$240\times360$                                     &RGB                            &472                &$\times$\\
\midrule
Ref. \cite{li2020video}                         &1                      &1                      &Intel Core i3-2130@3.4GHz                          &4GB                       &$352\times288$                         &YUV                            &5960$^\dag$        &$\times$\\
\midrule
Ref. \cite{yasser2020chaotic}                   &1                      &1                      &Intel Core i5-2400@3.1GHz                          &4GB                       &--                                     &RGB                            &408                &$\times$\\
\midrule
Ref. \cite{hosny2022fast}                       &1                      &1                      &Intel Core i7-8750H@2.2GHz                         &16GB                      &$352\times288$                         &RGB                            &260                &$\times$\\
\midrule
Ref. \cite{dua20223d}                           &1                      &1                      &Intel Core i5-6200U@2.3GHz                         &8GB                       &$360\times240$                         &RGB                            &5497               &$\times$\\
\midrule
\multirow{7}{1cm}{Ours}                         &\multirow{7}{0.5cm}{5} &\multirow{7}{0.5cm}{5} &\multirow{2}{4.6cm}{Intel Core i5-1135G7@2.4GHz}   &\multirow{2}{1.2cm}{8GB}  &\multirow{2}{1.6cm}{$480\times480$}    &\multirow{2}{1cm}{RGB}         &36.26 (PLCM)       &\checkmark\\
												&                       &                       &                                                   &                          &                                       &                               &35.49 (2DLASM)     &\checkmark\\
\cmidrule(r){4-9}
												&                       &                       &\multirow{2}{4.6cm}{Intel Core i7-8700@3.2GHz}     &\multirow{2}{1.2cm}{32GB} &\multirow{2}{1.6cm}{$576\times576$}    &\multirow{2}{1cm}{RGB}         &39.92 (PLCM)       &\checkmark\\
												&                       &                       &                                                   &                          &                                       &                               &40.67 (2DLASM)     &\checkmark\\
\cmidrule(r){4-9}
												&                       &                       &\multirow{2}{4.6cm}{Intel Xeon Gold 6226R@2.9GHz}  &\multirow{2}{1.2cm}{64GB} &\multirow{2}{1.6cm}{$768\times768$}    &\multirow{2}{1cm}{RGB}         &36.56 (PLCM)       &\checkmark\\
												&                       &                       &                                                   &                          &                                       &                               &36.23 (2DLASM)     &\checkmark\\
\midrule
\multicolumn{9}{l}{$r_c$: rounds of confusion; $r_d$: rounds of diffusion; CS: Color Space; AET: Average Encryption Time; RTE: Real-Time Encryption (AET}\\
\multicolumn{9}{l}{$\leq$ 1000 / FPS); --: not specified; $\dag$: the accurate average encryption time is not described in Ref. \cite{li2020video}, but is given in Ref. \cite{hosny2022fast}.}\\
\bottomrule
\label{Tab:comparison}
\end{tabular}
\end{table*}

\section{Comparison To Previous Works}
As discussed above, existing algorithms can be divided into full and selective video encryption. The proposed strategy, clearly, belongs to the former category. In this section, therefore, the deployed cryptosystems are compared with several recent published papers on full video encryption. The results are shown in Tab. \ref{Tab:comparison}. For some selected works, videos of different sizes are used to evaluate the encryption speed. We list the encryption speed for the largest videos in the table.

With the development of information science and hardware, the users put forward higher requirements for security. Thus almost all the latest works are based on confusion-diffusion architecture. To the best of our knowledge, however, these works only consist of one round of confusion and diffusion operation to improve the computational efficiency. And they cannot meet the requirements of real-time encryption, that is, the average encryption time (ms) of video frames is less than $1000~/$ FPS. Therefore, compare to previous works, the proposed strategy achieves the following advantages:

\begin{enumerate}[i)]
\item It realizes real-time video encryption, provides a feasible solution for related applications, and points out a new path for related research.
\item By taking advantages of parallel computing technique, it proves that real-time video encryption based on multi-round confusion-diffusion architecture is possible, and thus improves the security of real-time video encryption to the level of image encryption.
\item It works with many confusion, diffusion methods, and different chaotic maps.
\item It can be easily implemented with both software and hardware such as ARM, FPGA, etc.
\end{enumerate}

\section{Conclusion}

To sum up, a parallel computing technique based real-time chaotic video encryption strategy is proposed in this paper. It uses multiple threads to simultaneously perform confusion and diffusion operations on corresponding subframes to improve the computational efficiency. To evaluate the performance of the strategy, two chaotic maps are selected to implement cryptosystems. The encryption speed evaluation prove that such strategy significantly improves the encryption speed, and realizes real-time video encryption with different hardware platforms. The statistical and security analysis show that the deployed cryptosystems have outstanding statistical properties and resistance to attacks, noise, and data loss. Compared with previous works, the proposed strategy, to the best of our knowledge, achieves the fastest encryption speed, realizes the first real-time video encryption based on multiple-round confusion-diffusion architecture. This paper provides a feasible solution and a new path for related applications and research, respectively.


\section*{CRediT authorship contribution statement}
\textbf{Dong Jiang:} Conceptualization, Methodology, Writing-original draft,  Software, Supervision. \textbf{Zhen Yuan:} Investigation, Methodology, Software . \textbf{Wen-xin Li}: Investigation, Software. \textbf{Liang-liang Lu:} Conceptualization,  Methodology, Writing - review \& editing, Supervision.

\section*{Declaration of competing interest}
The authors declare that they have no known competing financial interests or personal relationships that could have appeared to influence the work reported in this paper. 

\section*{Data availability}
Data will be made available on request.

\section*{Acknowledgments}
This work is financially supported by the National Key Research and Development Program of China (2017YFA0303 704), the National Natural Science Foundation of China (12274233, 62273001), the Major Scientific Research Project of Anhui Province (KJ2021ZD0005),  the Compiled Scientific Research Plan Project of Anhui Province (2022AH040020), the Outstanding Research and Innovation Team Project of Anhui Province (2022AH010005), the University Collaborative Innovation Project of Anhui Province (GXXT-2021-091).






\end{document}